\documentclass[%
print,
superscriptaddress,
 amsmath,amssymb,
 aps,
]{revtex4}
\usepackage{soul}
\usepackage{float}
\usepackage[sc]{mathpazo}
\usepackage{graphicx}
\usepackage{dcolumn}
\usepackage{bm}
\usepackage{dsfont}
\usepackage[landscape,papersize={297.1mm,210mm},left=1.9cm,right=1.6cm,top=2.7cm,bottom=2.8cm]{geometry}
\usepackage[                   
            pdfstartview=FitH,
            colorlinks, 
            pdfborder=001,   
            linkcolor=blue,
            anchorcolor=blue,
            citecolor=blue,
            urlcolor=blue
            ]{hyperref}
\usepackage{amsthm}
\usepackage{multirow}
\usepackage{makecell}
\usepackage{booktabs,graphicx}
\makeatletter

\newcommand{\Rmnum}[1]{\expandafter\@slowromancap\romannumeral #1@}
\makeatother

\begin{document}
\title{Strong quantum nonlocality with genuine entanglement in an $N$-qutrit system}
\author{Mengying Hu}
\author{Ting Gao}
\email{gaoting@hebtu.edu.cn}
\affiliation{School of Mathematical Sciences, Hebei Normal University, Shijiazhuang 050024, China}
\affiliation{Hebei Mathematics Research Center,  Hebei Normal University, Shijiazhuang 050024, China}
\affiliation{Hebei International Joint Research Center for Mathematics and Interdisciplinary Science, Hebei Normal University, Shijiazhuang 050024, China}
\author{Fengli Yan}
\email{flyan@hebtu.edu.cn}
\affiliation{College of Physics, Hebei Key Laboratory of Photophysics Research and Application, Hebei Normal University, Shijiazhuang 050024, China}

\begin{abstract}
In this paper, we construct genuinely multipartite entangled bases in $(\mathbb{C}^{3})^{\otimes N}$ for $N\geq3$, where every state is one-uniform state. By modifying this construction, we successfully obtain strongly nonlocal orthogonal genuinely entangled sets and strongly nonlocal orthogonal genuinely entangled bases, which provide an answer to the open problem raised by Halder $et~al.$ [\href{https://journals.aps.org/prl/abstract/10.1103/PhysRevLett.122.040403} {Phy. Rev. Lett. \textbf{122}, 040403 (2019)}]. The strongly nonlocal orthogonal genuine entangled set we constructed in $(\mathbb{C}^{3})^{\otimes N}$ contains much fewer quantum states than all known ones. When $N>3$, our result answers the open question given by Wang $et~al$. [\href{https://journals.aps.org/pra/abstract/10.1103/PhysRevA.104.012424} {Phys. Rev. A \textbf{104}, 012424 (2021)}].

\textit{Keywords}: {strong quantum nonlocality; genuine entanglement; qutrit system}
\end{abstract}

\pacs{ 03.67.Mn, 03.65.Ud, 03.67.-a}

\maketitle

\section{Introduction}
Quantum nonlocality is a fundamental property of quantum mechanics, manifesting the nonclassical aspects of quantum phenomena. The most well-known manifestation of quantum nonlocality is Bell nonlocality \cite{N. Brunner}, which arises from entangled states \cite{R. Horodecki,T. Gao}. Entangled states show nonlocality by violating Bell-type inequalities \cite{J. S. Bell,J. F. Clauser,S. J. Freedman,A. Aspect,Z. Q. Chen,F. L. Yan,D. Ding1,H. X. Meng,D. Ding2}. It is well known that entanglement is an important resource in areas like quantum teleportation \cite{C. H. Bennett1,T. Gao2,T. Gao3}, quantum key distribution \cite{A. K. Ekert,H. K. Lo,C. H. Bennett2}, and quantum networks \cite{S. Perseguers}. On the other hand, the local indistinguishability of quantum states exhibits nonlocal properties in a way fundamentally different from Bell nonlocality. Local indistinguishability means that a known set of orthogonal quantum states distributed among spatially separated parties is not possible to be exactly distinguished by local operations and classical communication (LOCC) \cite{E. Chitambar}. In 1999, Bennett $et~al$. \cite{C. H. Bennett} presented a locally indistinguishable orthogonal product basis (OPB) in the Hilbert space $\mathbb{C}^{3}\otimes \mathbb{C}^{3}$, which shows the phenomenon of nonlocality without entanglement. Then, locally indistinguishable orthogonal product sets (OPSs) and orthogonal entangled sets (OESs) aroused widely research \cite{J. Walgate1,N. Yu1,J. Niset,Y. Feng,Z. C. Zhang1,Y. L. Wang1,Y. L. Wang2,S. Halder1,S. Ghosh1,S. Ghosh2,H. Fan,N. Yu2}, and found useful applications in data hiding \cite{B. M. Terha,D. P. DiVincenzo} and quantum secret sharing \cite{R. Rahaman,J. Wang}.

In 2019, Halder $et~al$. \cite{S. Halder2} introduced a stronger form of nonlocality, strong nonlocality, by the notion of local irreducibility of multipartite quantum states under every bipartition. Local irreducibility refers to that a set of orthogonal states cannot be eliminated one or more states from the set by orthogonal-preserving local measurements (OPLMs) \cite{S. Halder2}. A set that is locally irreducible must be locally indistinguishable, but the converse does not hold in general.
As OPSs in $\mathbb{C}^{2}\otimes\mathbb{C}^{d}$, $d\geq2$ are locally distinguishable, the locally irreducible phenomenon of OPS does not exist in the systems where one of the subsystems has a two-dimensional complex space. So Halder $et~al$. claimed that the multiparty OPS with strong nonlocality can only exist, if at all, on $H=\otimes_{i=1}^{N}H_{i}$, $N\geq3$, where dim$H_{i}\geq3$ for every $i$. And they provided two examples of strongly nonlocal OPB in $\mathbb{C}^{3}\otimes\mathbb{C}^{3}\otimes\mathbb{C}^{3}$ and $\mathbb{C}^{4}\otimes\mathbb{C}^{4}\otimes\mathbb{C}^{4}$. Zhang $et~al$. \cite{Z. C. Zhang2} presented a general definition of strong nonlocality for multipartite quantum systems, and distinguished the nonlocality of two sets of orthogonal quantum states. Later, strong quantum nonlocality without entanglement has been widely studied and many results are obtained. \cite{S. Rout,P. Yuan,F. Shi2,B. C. Che,H. Q. Zhou1,Y. He,H. Q. Zhou2}.

For genuinely entangled orthogonal bases (in which each element is entangled in every bipartition), intuition suggests that they are easier to exhibit strong nonlocality. However, Halder $et~al$. \cite{S. Halder2} found that the three-qubit Greenberger-Horne-Zeilinger (GHZ) basis (unnormalized) $\{|000\rangle\pm|111\rangle,~|011\rangle\pm|100\rangle,~|001\rangle\pm|110\rangle,~|010\rangle\pm|101\rangle\}$, which is genuinely entangled and locally irreducible (when all parts are separated), is locally reducible in all bipartitions. Then they asked whether one can find entangled bases that possess strong nonlocality.
In Ref. \cite{F. Shi3}, the authors showed strongly nonlocal OESs and strongly nonlocal orthogonal entangled bases (OEBs) in $\mathbb{C}^{d}\otimes\mathbb{C}^{d}\otimes\mathbb{C}^{d}~(d\geq3)$. However these states are not genuinely entangled. Wang $et~al$. \cite{Y. L. Wang} presented strongly nonlocal orthogonal genuinely entangled sets (OGESs) in $\mathbb{C}^{d}\otimes\mathbb{C}^{d}\otimes\mathbb{C}^{d}$ by using graph connectivity.
For multipartite quantum system, the authors of Ref. \cite{F. Shi1} provided strongly nonlocal OESs in ($\mathbb{C}^{d})^{\otimes N}$ for all $N\geq3$ and $d\geq2$, and strongly nonlocal OGESs when $N=3$ and $4$, but they did not present strongly nonlocal OGESs for $N\geq5$. Li $et~al$. \cite{M. S. Li1} constructed a strongly nonlocal OGES of size $\prod\limits_{n=1}^{N}d_{n}-\prod\limits_{n=1}^{N}(d_{n}-1)+1$ in multipartite quantum systems $H=\otimes^{N}_{n=1}H_{n}$ with $N\geq3$, where $d_{n}$ is the dimension of the $n$-th local subsystem $H_{n}$.

In this paper, we construct strongly nonlocal OGESs of size $2\times3^{N-1}$ and a strongly nonlocal orthogonal genuinely entangled basis (OGEB) in ($\mathbb{C}^{3})^{\otimes N}~(N\geq3)$. As a consequence, our constructions answer the open question: whether one can find entangled bases that are locally irreducible in all bipartitions, given by Halder $et~al$. [\href{https://journals.aps.org/prl/abstract/10.1103/PhysRevLett.122.040403} {Phy. Rev. Lett. \textbf{122}, 040403 (2019)}]. In an $N$-qutrit system, the OGES in our construction has $3^{N-1}-2^{N}+1$ states fewer than that constructed in Ref. \cite{M. S. Li1}.
In $\otimes^{N}_{i=1}\mathbb{C}^{d_{i}}~(N,d_{i}\geq3)$, the authors of Refs. \cite{Y. He} and \cite{H. Q. Zhou2} constructed strongly nonlocal OPSs with odd $N$ and even $N$ respectively, and both of them have size $3^{N}-1$ when $d_{i}=3$.
Ref. \cite{Y. He} also provided strongly nonlocal unextendible product bases (UPBs) of size $3^{N}-2^{N}$ in  $(\mathbb{C}^{3})^{\otimes N}$ with odd $N$. An UPB is a set of orthogonal product states which span a subspace of a given Hilbert space while the complementary subspace contains no product state \cite{P. Bej1}. In the same system, compared with these OPSs, the size $2\times3^{N-1}$ of OGESs in our construction is much smaller. Note that, in a $\mathbb{C}^{3}\otimes\mathbb{C}^{3}\otimes\mathbb{C}^{3}$ system, Shi $et~al.$ \cite{F. Shi1} showed a strongly nonlocal OGES with $18$ elements, which is consistent with our size. Che $et~al.$ \cite{B. C. Che} also constructed a UPB of size $12$. Our results provide an answer to the open problem in Ref.\cite{Y. L. Wang}, "Can we construct some smaller set that has the property of the strongest nonlocality via the OGES than the OPS."

The rest of this paper is organized as follows. In Sec. \ref{Q1}, we introduce some necessary notations and definitions used in the sequel. In Sec. \ref{Q2}, we construct an OGEB in $(\mathbb{C}^{3})^{\otimes N}~(N\geq3)$. In Sec \ref{Q3}, we exhibit strongly nonlocal OGESs and strongly nonlocal OGEBs in space $(\mathbb{C}^{3})^{\otimes N}~(N\geq3)$. Finally, we end with conclusions in Sec. \ref{Q4}.

\section{Preliminaries}\label{Q1}
Throughout this paper, we consider only pure state and do not normalize the states for simplicity.
For a $d$-dimensional Hilbert space $\mathbb{C}^{d}$ ($d\geq2$), we assume that $\mathcal{B}:=\{|0\rangle,|1\rangle,\cdots,|d-1\rangle\}$ is the computational basis of $\mathbb{C}^{d}$, and $\mathbb{Z}_{d} := \{0,1,\cdots,d-1\}$, $\mathbb{Z}^{N}_{d}:=(\mathbb{Z}_{d})^{\times N}$.
Given a $d\times d$ matrix $E:=\sum^{d-1}_{i=0}\sum^{d-1}_{j=0}a_{i,j}|i\rangle\langle j|,$ for $\mathcal{S}$, $\mathcal{T}\subseteq\{|0\rangle,|1\rangle,\cdots,|d-1\rangle\}$, we define
\begin{equation*}
_{\mathcal{S}}E_{\mathcal{T}}:=\sum\limits_{|s\rangle\in \mathcal{S}}\sum\limits_{|t\rangle\in \mathcal{T}}a_{s,t}|s\rangle\langle t|,
\end{equation*}
which is a sub-matrix of $E$ with row coordinates $\mathcal{S}$ and column coordinates $\mathcal{T}$. Especially, $_{\mathcal{S}}E_{\mathcal{S}}$ is represented by $E_{\mathcal{S}}$. A positive operator-valued measure (POVM) is a set of semidefinite operators $\{E_{m}=M^{\dagger}_{m}M_{m}\}$ such that $\sum_{m}E_{m}=\mathbb{I}$, where $\mathbb{I}$ is identity operation. A measurement is trivial if all the POVM elements are proportional to the identity operator, otherwise, the measurement is nontrivial \cite{J. Walgate2}. Clearly, the trivial measurement means that no information about the state can be yielded.

In a multipartite system $H_{A_{1}}\otimes\cdots \otimes H_{A_{N}}$ with a local dimension $d$ each, we say that $|\Psi\rangle_{A_{1}A_{2}\cdots A_{N}}$ is a one-uniform state \cite{D. Goyeneche} if its reduced density matrices for each subsystem is maximally mixed, i.e., $\rho_{A_{i}}=\mathrm{Tr}_{\bar{A}_{i}}(|\psi\rangle\langle\psi|)= \mathrm{I}/d$. A well-known example is $N$-qudit GHZ state $|\mathrm{GHZ}^{d}_{N}\rangle=\sum^{d-1}_{i=0}|i\rangle^{\otimes N}$.

Now we restate the definition of locally irreducible set and strong nonlocality \cite{S. Halder2}.

$\mathbf{Definition~1}$ (Locally irreducible set) A set $\{|\psi\rangle\}$ of orthogonal quantum states in  $\mathcal{H}=\otimes^{N}_{i=1}\mathbb{C}^{d_{i}}$ with $N\geq2$ and $d_{i}\geq2$, $i=1, 2, \cdots, N$, is locally irreducible if it is  not possible to eliminate one or more states from the set by OPLM.

$\mathbf{Definition~2}$ (Strong nonlocality) A set $\{|\psi\rangle\}$ of orthogonal quantum states in multipartite systems $\mathcal{H}=\otimes^{N}_{i=1}\mathbb{C}^{d_{i}}$ with $N\geq3$ and $d_{i}\geq2$, $i=1, 2, \cdots, N$, has the property of strong nonlocality if it is locally irreducible in every bipartition.

There is a sufficient condition for local irreducibility: if any parties can only perform a trivial OPLM, then the set of states must be locally irreducible. Therefore, one can show that a set $\{|\psi\rangle\}$ of orthogonal states is strongly nonlocal by proving that each subsystem of any bipartition can only perform a trivial OPLM.

Next, we state three lemmas of Shi $et~al$. \cite{F. Shi2} as follows.

$\mathbf{Lemma~1}$ (Block Zeros Lemma) Let an $n\times n$ matrix $E=(a_{i,j})_{i,j\in\mathbb{Z}_{n}}$ be matrix representation of the operator $E=M^{\dag}M$ under the bases $\mathcal{B}:=\{|0\rangle,|1\rangle,\cdots,|n-1\rangle\}$. Given two nonempty disjoint subsets $\mathcal{S}$ and $\mathcal{T}$ of $\mathcal{B}$, assume that $\{|\psi_{i}\rangle\}^{s-1}_{i=0}$, $\{|\phi_{j}\rangle\}^{t-1}_{j=0}$ are two orthogonal sets spanned by $\mathcal{S}$ and $\mathcal{T}$ respectively, where $s=|\mathcal{S}|$, $t=|\mathcal{T}|$. For $_{\mathcal{S}}E_{\mathcal{T}}:=\sum\limits_{|i\rangle\in \mathcal{S}}\sum\limits_{|j\rangle\in\mathcal{T}}a_{i,j}|i\rangle\langle j|$, if $\langle\psi_{i}|E|\phi_{j}\rangle=0$ for any $i\in\mathbb{Z}_{s}$, $j\in\mathbb{Z}_{t}$, then $_{\mathcal{S}}E_{\mathcal{T}}=\mathbf{0}$ and $_{\mathcal{T}}E_{\mathcal{S}}=\mathbf{0}$.

$\mathbf{Lemma~2}$ (Block Trivial Lemma) Let an $n\times n$ matrix $E=(a_{i,j})_{i,j\in\mathbb{Z}_{n}}$ be matrix representation of the operator $E=M^{\dag}M$ under the basis $\mathcal{B}:=\{|0\rangle,|1\rangle,\cdots,|n-1\rangle\}$. Given a nonempty subset $\mathcal{S}$ of $\mathcal{B}$, assume that $\{|\psi_{i}\rangle\}^{s-1}_{i=0}$ be an orthogonal set spanned by $\mathcal{S}$. Suppose that  $\langle\psi_{i}|E|\psi_{j}\rangle=0$ for any $i\neq j\in \mathbb{Z}_{s}$. If there exists a state $|u_{0}\rangle$, such that $_{\{|u_{0}\rangle\}}E_{S\setminus\{|u_{0}\rangle\}}=0$ and $\langle u_{0}|\psi_{j}\rangle\neq0$ for any $j\in \mathbb{Z}_{s}$, then $E_{\mathcal{S}}\propto \mathbb{I}_{\mathcal{S}}$.

$\mathbf{Lemma~3}$ Let $\{|\psi_{j}\rangle\}$ be a set of orthogonal states in multipartite system $\otimes^{N}_{i=1}\mathbb{C}^{d_{i}}$. For each $i=1,2,\cdots,N$, define $\bar{A}_{i}=\{A_{1}A_{2}\cdots A_{N}\}\backslash\{A_{i}\}$ is the joint party of all but the $i$-th party. If any OPLM on $\bar{A}_{i}$ is trivial, then the set $\{|\psi_{j}\rangle\}$ is of the strong nonlocality.

\section{THE CONSTRUCTION OF OGEBS in $(\mathbb{C}^{3})^{\otimes N}$}\label{Q2}
In this section, we construct three sets $\mathcal{G}^{N}_{i}$ ($i\in\mathbb{Z}_{3})$ of strings on $\mathbb{Z}^{N}_{3}$ and give two propositions to characterize them. Then we exhibit OGEBs in $(\mathbb{C}^{3})^{\otimes N}~(N\geq3)$ shown in Theorem~1.

Given a set $\mathcal{G}$ of $n$-tuples, we use the following method to obtain the $(n+1)$-tuples.
\begin{equation}
\begin{aligned}
\{j_{0}\}\times\mathcal{G}=&\{j_{0}\}\times\{(j^{1}_{1},\cdots,j^{1}_{n-1},j^{1}_{n}),(j^{2}_{1},\cdots,j^{2}_{n-1},j^{2}_{n}),\cdots, (j^{t}_{1},\cdots,j^{t}_{n-1},j^{t}_{n})\}\\
=&\{(j_{0},j^{1}_{1},\cdots,j^{1}_{n-1},j^{1}_{n}),(j_{0},j^{2}_{1},\cdots,j^{2}_{n-1},j^{2}_{n}),\cdots, (j_{0},j^{t}_{1},\cdots,j^{t}_{n-1},j^{t}_{n})\},
\end{aligned}
\end{equation}
where $t=|\mathcal{G}|$.

\subsection{THREE SETS on $\mathbb{Z}^{N}_{3}$}
\begin{figure}[h]
\centering
\includegraphics[height=40mm]{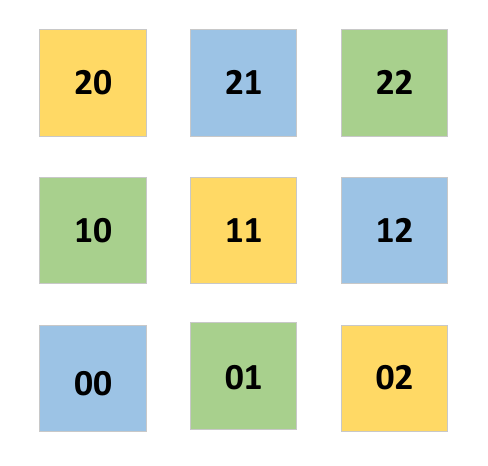}
\caption{The set $\mathbb{Z}_{3}\times\mathbb{Z}_{3}$ is depicted by a $3\times3$ grid, where the blue, green and yellow regions correspond to the set $\mathcal{G}^{2}_{0}$, $\mathcal{G}^{2}_{1}$ and $\mathcal{G}^{2}_{2}$ in Eq.(\ref{eq:G2}), respectively.}\label{fig:1}
\end{figure}

First, we consider $\mathbb{Z}_{3}$, and denote
\begin{equation}\label{eq:G1}
\mathcal{G}^{1}_{0}=\{0\},~~\mathcal{G}^{1}_{1}=\{1\},~~\mathcal{G}^{1}_{2}=\{2\}.
\end{equation}
For $N=2$, we give three subsets of $\mathbb{Z}_{3}\times \mathbb{Z}_{3}$,
\begin{equation}\label{eq:G2}
\begin{aligned}
&\mathcal{G}^{2}_{0}=\left(\{0\}\times \mathcal{G}^{1}_{0}\right) \bigcup \left(\{2\}\times \mathcal{G}^{1}_{1}\right) \bigcup \left(\{1\}\times \mathcal{G}^{1}_{2}\right)=\{(0,0), (2,1), (1,2)\},\\
&\mathcal{G}^{2}_{1}=\left(\{1\}\times \mathcal{G}^{1}_{0}\right) \bigcup \left(\{0\}\times \mathcal{G}^{1}_{1}\right) \bigcup \left(\{2\}\times \mathcal{G}^{1}_{2}\right)=\{(1,0), (0,1), (2,2)\},\\
&\mathcal{G}^{2}_{2}=\left(\{2\}\times \mathcal{G}^{1}_{0}\right) \bigcup \left(\{1\}\times \mathcal{G}^{1}_{1}\right) \bigcup \left(\{0\}\times \mathcal{G}^{1}_{2}\right)=\{(2,0), (1,1), (0,2)\}.
\end{aligned}
\end{equation}
Obviously, the sets $\mathcal{G}^{2}_{i}$ are pairwise disjoint and the union of all sets is $\mathbb{Z}_{3}\times \mathbb{Z}_{3}$ (see Fig. \ref{fig:1} for a visual exhibition).

Now, we construct three subsets of $\mathbb{Z}^{N}_{3}$ for $N\geq2$,
\begin{equation}\label{eq:G}
\begin{aligned}
&\mathcal{G}^{N}_{i}=\left(\{i\}\times \mathcal{G}^{N-1}_{0}\right) \bigcup \left(\{i\oplus_{3}2\}\times \mathcal{G}^{N-1}_{1}\right) \bigcup\left(\{i\oplus_{3}1\}\times \mathcal{G}^{N-1}_{2}\right),\\
\end{aligned}
\end{equation}
 where $i\in\mathbb{Z}_{3}$, $i\oplus_{3}t=(i+t)$ mod $3$. For each $i$, one can also exhibit the exact description
\begin{equation}\label{eq:G3}
\begin{aligned}
&\mathcal{G}^{N}_{0}=\left(\{0\}\times \mathcal{G}^{N-1}_{0}\right) \bigcup \left(\{2\}\times \mathcal{G}^{N-1}_{1}\right) \bigcup\left(\{1\}\times \mathcal{G}^{N-1}_{2}\right),\\
&\mathcal{G}^{N}_{1}=\left(\{1\}\times \mathcal{G}^{N-1}_{0}\right) \bigcup \left(\{0\}\times \mathcal{G}^{N-1}_{1}\right) \bigcup\left(\{2\}\times \mathcal{G}^{N-1}_{2}\right),\\
&\mathcal{G}^{N}_{2}=\left(\{2\}\times \mathcal{G}^{N-1}_{0}\right) \bigcup \left(\{1\}\times \mathcal{G}^{N-1}_{1}\right) \bigcup\left(\{0\}\times \mathcal{G}^{N-1}_{2}\right).\\
\end{aligned}
\end{equation}
Then, we have two propositions.

$\mathbf{Proposition~1}$ The sets given by Eq. (\ref{eq:G}) are pairwise disjoint and the union of all sets is $\mathbb{Z}^{N}_{3}$, that is,
\begin{equation*}
\mathcal{G}^{N}_{0}\cup\mathcal{G}^{N}_{1}\cup\mathcal{G}^{N}_{2}=\mathbb{Z}^{N}_{3}~\mathrm{and}~
\mathcal{G}^{N}_{i}\cap\mathcal{G}^{N}_{j}=\emptyset, ~\mathrm{where}~i\neq j\in\mathbb{Z}_{3}.
\end{equation*}

$\mathbf{Proof}$. According to Eq. (\ref{eq:G2}), it is clear that the claim is true for $N=2$.

We proceed by induction. Assume that the result has been proved for $N=k$, i.e., $\mathcal{G}^{k}_{0}\cup\mathcal{G}^{k}_{1}\cup\mathcal{G}^{k}_{2}$ = $\mathbb{Z}^{k}_{3}$ and $\mathcal{G}^{k}_{i}\cap\mathcal{G}^{k}_{j}=\emptyset$, $i\neq j\in\mathbb{Z}_{3}$. Let $l=k+1$, one gets
\begin{equation*}
\begin{aligned}
\mathcal{G}^{l}_{0}\cup\mathcal{G}^{l}_{1}\cup\mathcal{G}^{l}_{2} &= \bigg(\big(\{0\}\times \mathcal{G}^{k}_{0}\big) \cup \big(\{2\}\times \mathcal{G}^{k}_{1}\big)\cup \big(\{1\}\times \mathcal{G}^{k}_{2}\big)\bigg)\\
&~~~~~~~\bigcup\bigg(\big(\{1\}\times \mathcal{G}^{k}_{0}\big) \cup \big(\{0\}\times \mathcal{G}^{k}_{1}\big) \cup\big(\{2\}\times \mathcal{G}^{k}_{2}\big)\bigg)\\
&~~~~~~~\bigcup\bigg(\big(\{2\}\times \mathcal{G}^{k}_{0}\big) \cup \big(\{1\}\times \mathcal{G}^{k}_{1}\big) \cup\big(\{0\}\times \mathcal{G}^{k}_{2}\big)\bigg)\\
&=\bigg(\{0,1,2\}\times\mathcal{G}^{k}_{0}\bigg)\bigcup\bigg(\{0,1,2\}\times\mathcal{G}^{k}_{1}\bigg)\bigcup\bigg(\{0,1,2\}\times\mathcal{G}^{k}_{2}\bigg)\\
&=\{0,1,2\}\times\big(\mathcal{G}^{k}_{0}\cup\mathcal{G}^{k}_{1}\cup\mathcal{G}^{k}_{2}\big)\\
&=\{0,1,2\}\times\mathbb{Z}^{k}_{3}\\
&=\mathbb{Z}^{l}_{3}.
\end{aligned}
\end{equation*}\\
By the induction hypothesis, $\mathcal{G}^{k}_{i}\cap\mathcal{G}^{k}_{j}=\emptyset$ is true for $k$. Note that
$$\begin{cases}
\big(\{0\}\times \mathcal{G}^{k}_{0}\big) \bigcap \big(\{1\}\times \mathcal{G}^{k}_{0}\big)=\emptyset,\\
\big(\{0\}\times \mathcal{G}^{k}_{0}\big) \bigcap \big(\{0\}\times \mathcal{G}^{k}_{1}\big)=\emptyset,\\
\big(\{0\}\times \mathcal{G}^{k}_{0}\big)\bigcap\big(\{2\}\times \mathcal{G}^{k}_{2}\big)=\emptyset,
\end{cases}$$
it follows from Eq. (\ref{eq:G3}) that
\begin{equation}\label{eq:Gk0}
\big(\{0\}\times \mathcal{G}^{k}_{0}\big)\bigcap \mathcal{G}^{l}_{1}= \emptyset.\\
\end{equation}
Similarly, there are
\begin{equation}\label{eq:Gk}
\begin{cases}
\big(\{2\}\times \mathcal{G}^{k}_{1}\big)\bigcap \mathcal{G}^{l}_{1}= \emptyset,\\
\big(\{1\}\times \mathcal{G}^{k}_{2}\big)\bigcap \mathcal{G}^{l}_{1}= \emptyset.
\end{cases}
\end{equation}
 Combining Eqs. (\ref{eq:Gk0}) and (\ref{eq:Gk}) with Eq. (\ref{eq:G3}) (when $N=l$, $i=1$), we get $\mathcal{G}^{l}_{0}\cap \mathcal{G}^{l}_{1}= \emptyset$. Similarly, $\mathcal{G}^{l}_{0}\cap \mathcal{G}^{l}_{2}=\emptyset$ and $\mathcal{G}^{l}_{1}\cap \mathcal{G}^{l}_{2}=\emptyset$ can be deduced. $\hfill\blacksquare$

$\mathbf{Proposition~2}$ The set $\mathcal{G}^{N}_{i}$ given by Eq. (\ref{eq:G}) is invariant under arbitrary permutation.

$\mathbf{Proof}$. First, we rewrite Eq. (\ref{eq:G3}) as the following form
\begin{equation}\label{eq:M0}
\renewcommand\arraystretch{1.5}
\left [
\begin{array}{c}
\mathcal{G}^{N}_{0}\\
\mathcal{G}^{N}_{1}\\
\mathcal{G}^{N}_{2}
\end{array}
\right ]
=
\left [
\renewcommand\arraystretch{1.2}
\setlength{\arraycolsep}{1pt}
\begin{array}{ccc}
\{0\}&\{2\}&\{1\}\\
\{1\}&\{0\}&\{2\}\\
\{2\}&\{1\}&\{0\}\\
\end{array}
\right]
\times
\left [
\renewcommand\arraystretch{1.5}
\begin{array}{ccc}{}
\mathcal{G}^{N-1}_{0}\\
\mathcal{G}^{N-1}_{1}\\
\mathcal{G}^{N-1}_{2}
\end{array}
\right ]
=
\left [
\renewcommand\arraystretch{1.5}
\begin{array}{ccc}{}
(\{0\}\times \mathcal{G}^{N-1}_{0})\bigcup(\{2\}\times \mathcal{G}^{N-1}_{1})\bigcup(\{1\}\times \mathcal{G}^{N-1}_{2})\\
(\{1\}\times \mathcal{G}^{N-1}_{0})\bigcup(\{0\}\times \mathcal{G}^{N-1}_{1})\bigcup(\{2\}\times \mathcal{G}^{N-1}_{2})\\
(\{2\}\times \mathcal{G}^{N-1}_{0})\bigcup(\{1\}\times \mathcal{G}^{N-1}_{1})\bigcup(\{0\}\times \mathcal{G}^{N-1}_{2})
\end{array}
\right ].
\end{equation}
Similar to matrix multiplication, we get
\begin{equation}\label{eq:M3}
\renewcommand\arraystretch{1.5}
\left [
\begin{array}{ccc}
\mathcal{G}^{N}_{0}&\mathcal{G}^{N}_{2}&\mathcal{G}^{N}_{1}\\
\mathcal{G}^{N}_{1}&\mathcal{G}^{N}_{0}&\mathcal{G}^{N}_{2}\\
\mathcal{G}^{N}_{2}&\mathcal{G}^{N}_{1}&\mathcal{G}^{N}_{0}
\end{array}
\right ]
=
\left [
\renewcommand\arraystretch{1.2}
\setlength{\arraycolsep}{1pt}
\begin{array}{ccc}
\{0\}&\{2\}&\{1\}\\
\{1\}&\{0\}&\{2\}\\
\{2\}&\{1\}&\{0\}\\
\end{array}
\right]
\times
\left [
\renewcommand\arraystretch{1.5}
\begin{array}{ccc}{}
\mathcal{G}^{N-1}_{0}&\mathcal{G}^{N-1}_{2}&\mathcal{G}^{N-1}_{1}\\
\mathcal{G}^{N-1}_{1}&\mathcal{G}^{N-1}_{0}&\mathcal{G}^{N-1}_{2}\\
\mathcal{G}^{N-1}_{2}&\mathcal{G}^{N-1}_{1}&\mathcal{G}^{N-1}_{0}
\end{array}
\right ],
\end{equation}
where the result of the right side of Eq. (\ref{eq:M3}) is
\begin{equation*}
\footnotesize{
\left[
\renewcommand\arraystretch{1.5}
\setlength{\arraycolsep}{1pt}
\begin{array}{ccc}
(\{0\}\times \mathcal{G}^{N-1}_{0})\bigcup(\{2\}\times \mathcal{G}^{N-1}_{1}) \bigcup(\{1\}\times \mathcal{G}^{N-1}_{2})
~~&(\{0\}\times \mathcal{G}^{N-1}_{2})\bigcup(\{2\}\times \mathcal{G}^{N-1}_{0}) \bigcup(\{1\}\times \mathcal{G}^{N-1}_{1})
~~&(\{0\}\times \mathcal{G}^{N-1}_{1}) \bigcup(\{2\}\times \mathcal{G}^{N-1}_{2}) \bigcup(\{1\}\times \mathcal{G}^{N-1}_{0})\\
(\{1\}\times \mathcal{G}^{N-1}_{0}) \bigcup (\{0\}\times \mathcal{G}^{N-1}_{1}) \bigcup(\{2\}\times \mathcal{G}^{N-1}_{2})
~~&(\{1\}\times \mathcal{G}^{N-1}_{2}) \bigcup (\{0\}\times \mathcal{G}^{N-1}_{0}) \bigcup(\{2\}\times \mathcal{G}^{N-1}_{1})
~~&(\{1\}\times \mathcal{G}^{N-1}_{1}) \bigcup (\{0\}\times \mathcal{G}^{N-1}_{2}) \bigcup(\{2\}\times \mathcal{G}^{N-1}_{0})\\
(\{2\}\times \mathcal{G}^{N-1}_{0}) \bigcup(\{1\}\times \mathcal{G}^{N-1}_{1}) \bigcup(\{0\}\times \mathcal{G}^{N-1}_{2})
~~&(\{2\}\times \mathcal{G}^{N-1}_{2}) \bigcup (\{1\}\times \mathcal{G}^{N-1}_{0}) \bigcup(\{0\}\times \mathcal{G}^{N-1}_{1})
~~&(\{2\}\times \mathcal{G}^{N-1}_{1}) \bigcup (\{1\}\times \mathcal{G}^{N-1}_{2}) \bigcup(\{0\}\times \mathcal{G}^{N-1}_{0})
\end{array}
\right ].
}
\end{equation*}
Repeating this argument reveals
\begin{equation}\label{eq:M1}
\left [
\renewcommand\arraystretch{1.5}
\begin{array}{ccc}
\mathcal{G}^{N}_{0}&\mathcal{G}^{N}_{2}&\mathcal{G}^{N}_{1}\\
\mathcal{G}^{N}_{1}&\mathcal{G}^{N}_{0}&\mathcal{G}^{N}_{2}\\
\mathcal{G}^{N}_{2}&\mathcal{G}^{N}_{1}&\mathcal{G}^{N}_{0}
\end{array}
\right ]_{[N,\cdots,2,1]}
=
\left [
\renewcommand\arraystretch{1.2}
\setlength{\arraycolsep}{1pt}
\begin{array}{ccc}
\{0\}&\{2\}&\{1\}\\
\{1\}&\{0\}&\{2\}\\
\{2\}&\{1\}&\{0\}
\end{array}
\right ]_{N}
\times
\cdots
\times
\left [
\renewcommand\arraystretch{1.2}
\setlength{\arraycolsep}{1pt}
\begin{array}{ccc}
\{0\}&\{2\}&\{1\}\\
\{1\}&\{0\}&\{2\}\\
\{2\}&\{1\}&\{0\}
\end{array}
\right]_{2}
\times
\left [
\renewcommand\arraystretch{1.3}
\setlength{\arraycolsep}{1.2pt}
\begin{array}{ccc}
\mathcal{G}^{1}_{0}&\mathcal{G}^{1}_{2}&\mathcal{G}^{1}_{1}\\
\mathcal{G}^{1}_{1}&\mathcal{G}^{1}_{0}&\mathcal{G}^{1}_{2}\\
\mathcal{G}^{1}_{2}&\mathcal{G}^{1}_{1}&\mathcal{G}^{1}_{0}
\end{array}
\right ]_{1}.
\end{equation}
 Suppose that the elements in $\mathcal{G}^{N}_{i}$ we constructed are ordered strings. For example, consider any string $(c_{N},\cdots,b_{x},\cdots,a_{1})_{[N,\cdots,x,\cdots,1]}$ belongs to $\mathcal{G}^{N}_{i}$, where $[N,\cdots,x,\cdots,1]$ indicates the position order of each element in this string, and the index $x$ means that the element $b_{x}$ comes from the $x$-th square matrix on the right side.

To prove $\mathcal{G}^{N}_{i}$ is invariant under arbitrary permutation, we only need to show that the elements in $\mathcal{G}^{N}_{i}$ will not change under arbitrary permutation of the position order $[N,\cdots,2,1]$. Substituting Eq. (\ref{eq:G1}) into Eq. (\ref{eq:M1}), we get
\begin{equation}\label{eq:M2}
\left [
\renewcommand\arraystretch{1.5}
\begin{array}{ccc}
\mathcal{G}^{N}_{0}&\mathcal{G}^{N}_{2}&\mathcal{G}^{N}_{1}\\
\mathcal{G}^{N}_{1}&\mathcal{G}^{N}_{0}&\mathcal{G}^{N}_{2}\\
\mathcal{G}^{N}_{2}&\mathcal{G}^{N}_{1}&\mathcal{G}^{N}_{0}
\end{array}
\right ]_{[N,\cdots,2,1]}
=
\left [
\renewcommand\arraystretch{1.2}
\setlength{\arraycolsep}{1pt}
\begin{array}{ccc}
\{0\}&\{2\}&\{1\}\\
\{1\}&\{0\}&\{2\}\\
\{2\}&\{1\}&\{0\}
\end{array}
\right ]_{N}
\times
\cdots
\times
\left [
\renewcommand\arraystretch{1.2}
\setlength{\arraycolsep}{1pt}
\begin{array}{ccc}
\{0\}&\{2\}&\{1\}\\
\{1\}&\{0\}&\{2\}\\
\{2\}&\{1\}&\{0\}
\end{array}
\right]_{2}
\times
\left [
\renewcommand\arraystretch{1.3}
\setlength{\arraycolsep}{1pt}
\begin{array}{ccc}
\{0\}&\{2\}&\{1\}\\
\{1\}&\{0\}&\{2\}\\
\{2\}&\{1\}&\{0\}
\end{array}
\right ]_{1}.
\end{equation}
Because the right square matrices are the same, let ${[i_{N},\cdots,i_{2},i_{1}]}$ be an arbitrary permutation of $[N,\cdots,2,1]$, we have
\begin{equation}
\left [
\renewcommand\arraystretch{1.5}
\begin{array}{ccc}
\mathcal{G}^{N}_{0}&\mathcal{G}^{N}_{2}&\mathcal{G}^{N}_{1}\\
\mathcal{G}^{N}_{1}&\mathcal{G}^{N}_{0}&\mathcal{G}^{N}_{2}\\
\mathcal{G}^{N}_{2}&\mathcal{G}^{N}_{1}&\mathcal{G}^{N}_{0}
\end{array}
\right ]_{[i_{N},\cdots,i_{2},i_{1}]}
=
\left [
\renewcommand\arraystretch{1.2}
\setlength{\arraycolsep}{1pt}
\begin{array}{ccc}
\{0\}&\{2\}&\{1\}\\
\{1\}&\{0\}&\{2\}\\
\{2\}&\{1\}&\{0\}
\end{array}
\right ]_{i_{N}}
\times
\cdots
\times
\left [
\renewcommand\arraystretch{1.2}
\setlength{\arraycolsep}{1pt}
\begin{array}{ccc}
\{0\}&\{2\}&\{1\}\\
\{1\}&\{0\}&\{2\}\\
\{2\}&\{1\}&\{0\}
\end{array}
\right]_{i_{2}}
\times
\left [
\renewcommand\arraystretch{1.3}
\setlength{\arraycolsep}{1pt}
\begin{array}{ccc}
\{0\}&\{2\}&\{1\}\\
\{1\}&\{0\}&\{2\}\\
\{2\}&\{1\}&\{0\}
\end{array}
\right ]_{i_{1}}.
\end{equation}
Therefore the proof is now complete. $\hfill\blacksquare$

\subsection{OGEBS IN $(\mathbb{C}^{3})^{\otimes N}$}

Let $\mathcal{H}:=(\mathbb{C}^{3})^{\otimes N}$, $s_{i}$ be the cardinality of the set $\mathcal{G}^{N}_{i}$ given by Eq. (\ref{eq:G}), $i\in \mathbb{Z}_{3}$. Define
\begin{equation}\label{eq:S}
\begin{aligned}
&\mathcal{S}_{i}:=\{|\Psi_{i,k}\rangle \in \mathcal{H}~\big|~k \in \mathbb{Z}_{s_{i}}, |\Psi_{i,k}\rangle:=\sum\limits_{\textbf{\emph{j}} \in {\mathcal{G}^{N}_{i}}}\omega^{k f_{i}(\textbf{\emph{j}})}_{s_{i}}|\textbf{\emph{j}}\rangle\}.\\
\end{aligned}
\end{equation}
 Here $ f_{i}:\mathcal{G}^{N}_{i}\longrightarrow \mathbb{Z}_{s_{i}}$ is any fixed bijection and $\omega_{s_{i}}:=\mathrm{e}^{\frac{2\pi\sqrt{-1}}{s_{i}}}$, $s_{i}=|\mathcal{G}^{N}_{i}|$. Evidently, the set $\{\mathcal{S}_{i}\}$ of states is an orthogonal basis in $(\mathbb{C}^{3})^{\otimes N}$. Furthermore, it forms a genuinely entangled orthogonal basis.

$\mathbf{Theorem~1}$ The set $\bigcup^{2}_{i=0}\mathcal{S}_{i}$ of states given by Eq. (\ref{eq:S}) is an OGEB in $(\mathbb{C}^{3})^{\otimes N}$.

$\mathbf{Proof.}$ we only need to show that $|\Psi_{i,k}\rangle$ is entangled for each bipartition of the system $\{\mathcal{A}_{1},\mathcal{A}_{2},\cdots,\mathcal{A}_{N}\}$. Let $\{\mathcal{A}_{x_{1}}\mathcal{A}_{x_{2}}\cdots\mathcal{A}_{x_{s}}\}|\{\mathcal{A}_{x_{s+1}}\mathcal{A}_{x_{s+2}}\cdots\mathcal{A}_{x_{N}}\}$ ($1\leq s\leq N-1$) is a bipartition of the subsystem, where $\{x_{1},x_{2},\cdots,x_{N}\}$ is an arbitrary permutation of $\{1,2,\cdots,N\}$.
We denote $\mathcal{A}$ and $\mathcal{B}$ as the computational bases of the systems $\{\mathcal{A}_{x_{1}}\mathcal{A}_{x_{2}}\cdots\mathcal{A}_{x_{s}}\}$ and $\{\mathcal{A}_{x_{s+1}}\mathcal{A}_{x_{s+2}}\cdots\mathcal{A}_{x_{N}}\}$ respectively, and express state $|\Psi_{i,k}\rangle$ as
\begin{equation*}
|\Psi_{i,k}\rangle=\Sigma_{|a\rangle\in\mathcal{A}}\Sigma_{|b\rangle\in\mathcal{B}}\psi_{a,b}|a\rangle|b\rangle.
\end{equation*}
Then, $|\Psi_{i,k}\rangle$ is entangled if the rank of the matrix $(\psi_{a,b})$ is greater than one.

Now, we state a fact obtained by Eq. (\ref{eq:G3}): for arbitrary $(i_{1},i_{2},\cdots,i_{N-1})$ $\in$ $\mathds{Z}^{N-1}_{3}$, the following strings $(0,i_{1},i_{2},\cdots,i_{N-1})$, $(1,i_{1},i_{2},\cdots,i_{N-1})$, $(2,i_{1},i_{2},\cdots,i_{N-1})$ are distributed in different sets.  As a consequence, given $(j_{1},j_{2},\cdots,j_{N-2})\in\mathcal{G}^{N-2}_{0}$, we get
\begin{equation}
\begin{cases}
(0,j_{1},j_{2},\cdots,j_{N-2})\in\mathcal{G}^{N-1}_{0},\\
(1,j_{1},j_{2},\cdots,j_{N-2})\in\mathcal{G}^{N-1}_{1},\\
(2,j_{1},j_{2},\cdots,j_{N-2})\in\mathcal{G}^{N-1}_{2}.\label{eq:differ}
\end{cases}
\end{equation}
By Proposition 2 we have
$$\begin{cases}
(j_{1},j_{2},\cdots,j_{N-2},0)\in\mathcal{G}^{N-1}_{0},\\
(j_{1},j_{2},\cdots,j_{N-2},1)\in\mathcal{G}^{N-1}_{1},\\
(j_{1},j_{2},\cdots,j_{N-2},2)\in\mathcal{G}^{N-1}_{2}.
\end{cases}$$
Then, it follows immediately from Eq. (\ref{eq:G}) that for any $i\in\mathbb{Z}_{3}$
$$\begin{cases}
(i,j_{1},j_{2},\cdots,j_{N-2},0)\in\mathcal{G}^{N}_{i},\\
(i\oplus_{3}2,j_{1},j_{2},\cdots,j_{N-2},1)\in\mathcal{G}^{N}_{i},\\
(i\oplus_{3}1,j_{1},j_{2},\cdots,j_{N-2},2)\in\mathcal{G}^{N}_{i},
\end{cases}$$
and
\begin{equation}
\begin{cases}
(i,j_{1},j_{2},\cdots,j_{N-2},1)\notin\mathcal{G}^{N}_{i},\\
(i,j_{1},j_{2},\cdots,j_{N-2},2)\notin\mathcal{G}^{N}_{i},\\
(i\oplus_{3}2,j_{1},j_{2},\cdots,j_{N-2},0)\notin\mathcal{G}^{N}_{i},\\
(i\oplus_{3}2,j_{1},j_{2},\cdots,j_{N-2},2)\notin\mathcal{G}^{N}_{i},\\
(i\oplus_{3}1,j_{1},j_{2},\cdots,j_{N-2},0)\notin\mathcal{G}^{N}_{i},\\
(i\oplus_{3}1,j_{1},j_{2},\cdots,j_{N-2},1)\notin\mathcal{G}^{N}_{i}.\label{eq:notin}
\end{cases}
\end{equation}
If the statement about Eq. (\ref{eq:notin}) was not true, then at least one of the above strings belongs to $\mathcal{G}^{N}_{i}$. Assume that $(i,j_{1},j_{2},\cdots,j_{N-2},1)\in\mathcal{G}^{N}_{i}$. Proposition 2 ensures that $(1,i,j_{1},j_{2},\cdots,j_{N-2})$ and $(0,i,j_{1},j_{2},\cdots,j_{N-2})$ belong to the same set $\mathcal{G}^{N}_{i}$. Evidently, this contradicts the fact we originally stated, thus $(i,j_{1},\cdots,j_{N-2},1)\notin\mathcal{G}^{N}_{i}$.  The other cases of Eq. (\ref{eq:notin}) can be proved similarly.

Based on the above argument, we can conclude that matrix $(\psi_{a,b})$ has one of the following two $2\times2$ submatrices,
\begin{table}[!htp]
\renewcommand\arraystretch{1.5}
\begin{tabular}{c c c c c c}
~~~&$|j_{s}\cdots j_{(N-2)}0\rangle$&~~~$|j_{s}\cdots j_{(N-2)}1\rangle$&~~~~~~~~&$|j_{s}\cdots j_{(N-2)}0\rangle$&~~~$|j_{s}\cdots j_{(N-2)}1\rangle$\\
$|i~j_{1}j_{2}\cdots j_{s-1}\rangle$ & $\alpha_{1}$ & $0$&~~~~~~~~~    $|(i\oplus_{3}2)j_{1}j_{2}\cdots j_{s-1}\rangle$&    $0$ &    $\beta_{2}$           \\
$|(i\oplus_{3}2)j_{1}j_{2}\cdots j_{s-1}\rangle$ &$0$& $\beta_{1}$&~~~$|i~j_{1}j_{2}\cdots j_{s-1}\rangle$&   $\alpha_{2}$ & $0$ \\
\end{tabular}
\end{table}\\
where $\alpha_{m}\beta_{m}\neq0~(m=1,2)$. So the Schmidt rank of $|\Psi_{i,k}\rangle$ under each bipartition is greater than one. Hence $|\Psi_{i,k}\rangle$ is a genuinely entangled state.

Similarly, if one chooses the string $(j_{1},j_{2},\cdots,j_{N-2})$ from $\mathcal{G}^{N-2}_{1}$ or $\mathcal{G}^{N-2}_{2}$, the same conclusion can be obtained.$\hfill\blacksquare$

Furthermore, each state in $\bigcup^{2}_{i=0}\mathcal{S}_{i}$ is one-uniform state. The reasons are as follows. For any string $(j_{1},j_{2},\cdots,j_{N-1})\in\mathds{Z}^{N-1}_{3}$, the base vector corresponding to $(0,j_{1},j_{2},\cdots,j_{N-1})$, $(1, j_{1},j_{2},\cdots,j_{N-1})$ and $(2,j_{1},j_{2},\cdots,j_{N-1})$ must belong to the superpositions of different states. It is easy to check that all of its one-qubit reductions are maximally mixed. The bases in an $N$-qudit systems is called a 'maximum entangled basis' (MEB) \cite{W. M. Shang} if each element is one-uniform state.  MEB has been found useful applications in quantum information masking. For example, the authors of Ref. \cite{W. M. Shang} used MEB and showed that it's possible to mask arbitrary unknown quantum states into multipartite lower-dimensional systems.

\section{Strongly nonlocal OGES AND strongly nonlocal OGEBs In $(\mathbb{C}^{3})^{\otimes N}$}\label{Q3}
In this section, by modifying the previous construction, we successfully show strongly nonlocal OGESs of size $2\times3^{N-1}$ and  strongly nonlocal OGEBs in Hilbert space $\mathcal{H}=(\mathbb{C}^{3})^{\otimes N}$. Our OGESs are strictly fewer, $3^{N-1}-2^{N}+1$ fewer to be precise, than the size $3^{N}-2^{N}+1$ of the strongly nonlocal OGES in Ref. \cite{M. S. Li1}. We prove that only $2\times3^{N-1}$ entangled states can also exhibit strong nonlocality in an $N$-qutrit system.

Let $(0)^{\times N}:=(\underbrace{0,0,\cdots,0}\limits_{N})$, $(1)^{\times N}:=(\underbrace{1,1,\cdots,1}\limits_{N})$, $(2)^{\times N}:=(\underbrace{2,2,\cdots,2}\limits_{N})$ and denote $\mathbf{0}=(0)^{\times (N-1)}$, $\mathbf{1}=(1)^{\times (N-1)}$, $\mathbf{2}=(2)^{\times (N-1)}$. Based on the distribution of $(0)^{\times N}$, $(1)^{\times N}$ and $(2)^{\times N}$ in sets $\mathcal{G}^{N}_{i}~(i\in\mathds{Z}_{3})$, our construction is discussed in three cases. For the detail distribution please see TABLE \ref{tab:1}.
\begin{table}[H]\centering
\caption{The distribution of $(0)^{\times N}$, $(1)^{\times N}$ and $(2)^{\times N}$, where $n\geq1$.}
\renewcommand\arraystretch{1.5}
\begin{tabular}{l c  c  c}
 \hline
 \hline
System $N$& $(0)^{\times N}$ & $(1)^{\times N}$ & $(2)^{\times N}$\\
 \hline
$N=2$ & $(0,0)\in\mathcal{G}^{2}_{0}$  & $(1,1)\in\mathcal{G}^{2}_{2}$ & $(2,2)\in\mathcal{G}^{2}_{1}$ \\

$N=3$ & $(0,0,0)\in\big(\{0\}\times\mathcal{G}^{2}_{0}\big)\subset\mathcal{G}^{3}_{0}$ &
$(1,1,1)\in\big(\{1\}\times\mathcal{G}^{2}_{2}\big)\subset\mathcal{G}^{3}_{0}$ &
$(2,2,2)\in\big(\{2\}\times\mathcal{G}^{2}_{1}\big)\subset\mathcal{G}^{3}_{0}$ \\

$N=4$ &$(0)^{\times4}\in\big(\{0\}\times\mathcal{G}^{3}_{0}\big)\subset\mathcal{G}^{4}_{0}$ & $(1)^{\times4}\in\big(\{1\}\times\mathcal{G}^{3}_{0}\big)\subset\mathcal{G}^{4}_{1}$ & $(2)^{\times4}\in\big(\{2\}\times\mathcal{G}^{3}_{0}\big)\subset\mathcal{G}^{4}_{2}$\\

$N=5$ &$(0)^{\times5}\in\big(\{0\}\times\mathcal{G}^{4}_{0}\big)\subset\mathcal{G}^{5}_{0}$ & $(1)^{\times5}\in\big(\{1\}\times\mathcal{G}^{4}_{1}\big)\subset\mathcal{G}^{5}_{2}$ & $(2)^{\times5}\in\big(\{2\}\times\mathcal{G}^{4}_{2}\big)\subset\mathcal{G}^{5}_{1}$\\

$N=6$ &$(0)^{\times6}\in\big(\{0\}\times\mathcal{G}^{5}_{0}\big)\subset\mathcal{G}^{6}_{0}$ & $(1)^{\times6}\in\big(\{1\}\times\mathcal{G}^{5}_{2}\big)\subset\mathcal{G}^{6}_{0}$ & $(2)^{\times6}\in\big(\{2\}\times\mathcal{G}^{5}_{1}\big)\subset\mathcal{G}^{6}_{0}$\\

$N=7$ &$(0)^{\times7}\in\big(\{0\}\times\mathcal{G}^{6}_{0}\big)\subset\mathcal{G}^{7}_{0}$ & $(1)^{\times7}\in\big(\{1\}\times\mathcal{G}^{6}_{0}\big)\subset\mathcal{G}^{7}_{1}$ & $(2)^{\times7}\in\big(\{2\}\times\mathcal{G}^{6}_{0}\big)\subset\mathcal{G}^{7}_{2}$\\

$N=8$ &$(0)^{\times8}\in\big(\{0\}\times\mathcal{G}^{7}_{0}\big)\subset\mathcal{G}^{8}_{0}$ & $(1)^{\times8}\in\big(\{1\}\times\mathcal{G}^{7}_{1}\big)\subset\mathcal{G}^{8}_{2}$ & $(2)^{\times8}\in\big(\{2\}\times\mathcal{G}^{7}_{2}\big)\subset\mathcal{G}^{8}_{1}$\\

$N=9$ &$(0)^{\times9}\in\big(\{0\}\times\mathcal{G}^{8}_{0}\big)\subset\mathcal{G}^{9}_{0}$ & $(1)^{\times9}\in\big(\{1\}\times\mathcal{G}^{8}_{2}\big)\subset\mathcal{G}^{9}_{0}$ & $(2)^{\times9}\in\big(\{2\}\times\mathcal{G}^{8}_{1}\big)\subset\mathcal{G}^{9}_{0}$\\

\vdots &\vdots & \vdots & \vdots\\

$N=3n$ &$(0)^{\times3n}\in\big(\{0\}\times\mathcal{G}^{(3n-1)}_{0}\big)\subset\mathcal{G}^{N}_{0}$ & $(1)^{\times3n}\in\big(\{1\}\times\mathcal{G}^{(3n-1)}_{2}\big)\subset\mathcal{G}^{N}_{0}$ & $(2)^{\times3n}\in\big(\{2\}\times\mathcal{G}^{(3n-1)}_{1}\big)\subset\mathcal{G}^{N}_{0}$\\

$N=3n+1$ &$(0)^{\times(3n+1)}\in\big(\{0\}\times\mathcal{G}^{3n}_{0}\big)\subset\mathcal{G}^{N}_{0}$ & $(1)^{\times(3n+1)}\in\big(\{1\}\times\mathcal{G}^{3n}_{0}\big)\subset\mathcal{G}^{N}_{1}$ & $(2)^{\times(3n+1)}\in\big(\{2\}\times\mathcal{G}^{3n}_{0}\big)\subset\mathcal{G}^{N}_{2}$\\

$N=3n+2$ &$(0)^{\times(3n+2)}\in\big(\{0\}\times\mathcal{G}^{(3n+1)}_{0}\big)\subset\mathcal{G}^{N}_{0}$ & $(1)^{\times(3n+2)}\in\big(\{1\}\times\mathcal{G}^{(3n+1)}_{1}\big)\subset\mathcal{G}^{N}_{2}$ &
$(2)^{\times(3n+2)}\in\big(\{2\}\times\mathcal{G}^{(3n+1)}_{2}\big)\subset\mathcal{G}^{N}_{1}$\\
\hline
\hline
\end{tabular}\label{tab:1}
\end{table}

$\mathbf{Case~I}$: When $N=3n~(n\geq1)$, we redefine
\begin{equation}
\begin{aligned}
&\widetilde{\mathcal{G}^{N}_{0}}=\left(\{0\}\times \big(\mathcal{G}^{N-1}_{0}\backslash\{\mathbf{0}\}\big)\right) \bigcup \left(\{2\}\times \big(\mathcal{G}^{N-1}_{1}\backslash\{\mathbf{2}\}\big)\right) \bigcup\left(\{1\}\times \big(\mathcal{G}^{N-1}_{2}\backslash\{\mathbf{1}\}\big)\right),\\
&\widetilde{\mathcal{G}^{N}_{1}}=\left(\{1\}\times \mathcal{G}^{N-1}_{0}\right) \bigcup \left(\{0\}\times \mathcal{G}^{N-1}_{1}\right) \bigcup\left(\{2\}\times \mathcal{G}^{N-1}_{2}\right),\\
&\widetilde{\mathcal{G}^{N}_{2}}=\left(\{2\}\times \mathcal{G}^{N-1}_{0}\right) \bigcup \left(\{1\}\times \mathcal{G}^{N-1}_{1}\right) \bigcup\left(\{0\}\times \mathcal{G}^{N-1}_{2}\right),\\
&\widetilde{\mathcal{G}^{N}_{3}}=\big\{(0,\underbrace{0,\cdots,0}\limits_{N-1}), (1,\underbrace{1,\cdots,1}\limits_{N-1}), (2,\underbrace{2,\cdots,2}\limits_{N-1})\big\}.
\end{aligned}\label{eq:case1G}
\end{equation}
Here $\mathcal{G}^{N-1}_{i}~(i=0,1,2)$ is given by Eq. (\ref{eq:G}). Clearly we just rename sets $\mathcal{G}^{N}_{1}$ and $\mathcal{G}^{N}_{2}$ that given by Eq. (\ref{eq:G3}) as $\widetilde{\mathcal{G}^{N}_{1}}$ and $\widetilde{\mathcal{G}^{N}_{2}}$  without changing their structure.

Let $\widetilde{s_{i}}$ be the cardinality of the set $\widetilde{\mathcal{G}^{N}_{i}}$, $i\in \mathbb{Z}_{4}$, we define
\begin{equation}
\begin{aligned}
&\widetilde{\mathcal{S}_{i}}:=\{|\widetilde{\Psi}_{i,k}\rangle \in \mathcal{H}~\big|~k \in \mathbb{Z}_{\widetilde{s_{i}}}, |\widetilde{\Psi}_{i,k}\rangle:=\sum\limits_{\textbf{\emph{j}}\in {\widetilde{\mathcal{G}^{N}_{i}}}}\omega^{k f_{i}(\textbf{\emph{j}})}_{\widetilde{s_{i}}}|\textbf{\emph{j}}\rangle\}.
\end{aligned}\label{eq:IVS1}
\end{equation}
Here $ f_{i}:\widetilde{\mathcal{G}^{N}_{i}}\longrightarrow \mathbb{Z}_{\widetilde{s_{i}}}$ is any fixed bijection and $\omega_{\widetilde{s_{i}}}:=\mathrm{e}^{\frac{2\pi\sqrt{-1}}{\widetilde{s_{i}}}}$.

Comparing Eq. (\ref{eq:case1G}) with Eq. (\ref{eq:G3}), we only change the position of $(0)^{\times N}$, $(2)^{\times N}$ and $(1)^{\times N}$.
Then according to the proof of Proposition 2 and Theorem 1, the change of these elements does not change the permutation invariance of sets $\widetilde{\mathcal{G}_{i}}$ in Eq. (\ref{eq:case1G}), nor does it change the genuine entanglement of sets $\widetilde{\mathcal{S}_{i}}$ in Eq. (\ref{eq:IVS1}).

$\mathbf{Theorem~2}$ In $(\mathbb{C}^{3})^{\otimes N}~(N=3n,~n\geq1)$, the set $\widetilde{\mathcal{S}}:=\cup^{3}_{i=0}\widetilde{\mathcal{S}_{i}}$ given by Eq. (\ref{eq:IVS1}) is a strongly nonlocal OGEB. The set $\widetilde{\mathcal{S}}\backslash\widetilde{\mathcal{S}_{2}}=\widetilde{\mathcal{S}_{0}}\cup\widetilde{\mathcal{S}_{1}}\cup\widetilde{\mathcal{S}_{3}}$ is a strongly nonlocal OGES of size $2\times3^{N-1}$.

$\mathbf{Proof}$. Evidently, $\widetilde{\mathcal{S}}$ is an OGEB in $(\mathbb{C}^{3})^{\otimes N}$. Thus, we only need to prove that $\widetilde{\mathcal{S}}\backslash\widetilde{\mathcal{S}_{2}}$ has the property of strong nonlocality. First, we show that $A_{2}A_{3}\cdots A_{N}$ can only perform a trivial OPLM $\{\Pi_{\alpha}\}$, $\alpha=1,2,\cdots$. Let $\Pi_{\alpha}=M^{\dagger}_{\alpha}M_{\alpha}$, since the measurement is orthogonality-preserving, for every $\alpha$,  the postmeasurement states must be pairwise orthogonal
\begin{equation*}
\langle\Psi|\mathbb{I}_{A_{1}}\otimes M^{\dagger}_{\alpha}M_{\alpha}|\Phi\rangle=\langle\Psi|\mathbb{I}_{A_{1}}\otimes \Pi_{\alpha}|\Phi\rangle=0
\end{equation*}
for any two different states $|\Psi\rangle, |\Phi\rangle\in \widetilde{\mathcal{S}}\backslash\widetilde{\mathcal{S}_{2}}$. According to Proposition~1, the sets of basis vectors corresponding to $\widetilde{\mathcal{G}^{N}_{0}}$, $\widetilde{\mathcal{G}^{N}_{1}}$ and $\widetilde{\mathcal{G}^{N}_{3}}$ are disjoint subsets of the computational basis  $\mathcal{B}=\{\otimes^{N}_{k=1}|\eta_{k}\rangle\big|\eta_{k}=0,1,2\}$ of $(\mathbb{C}^{3})^{\bigotimes N}$.

 Applying Block Zeros Lemma to any two different sets $\widetilde{\mathcal{S}_{i}}$ and $\widetilde{\mathcal{S}_{j}}$ $(i\neq j\in\{0,1,3\})$, we obtain
\begin{equation*}
\langle i_{1},i_{2},\cdots, i_{N}|\mathbb{I}_{A_{1}}\otimes \Pi_{\alpha}|j_{1},j_{2},\cdots,j_{N}\rangle=\langle\textbf{\emph{i}}|E|\textbf{\emph{j}}\rangle=0\\
\end{equation*}
for any $\textbf{\emph{i}}=(i_{1},i_{2},\cdots, i_{N})\in \widetilde{\mathcal{G}^{N}_{i}}$ and $\textbf{\emph{j}}=(j_{1},j_{2},\cdots,j_{N})\in\widetilde{\mathcal{G}^{N}_{j}}$. Here $E:=\mathbb{I}_{A_{1}}\otimes \Pi_{\alpha}$.

When $i_{1}=j_{1}$, one gets
\begin{equation*}
\langle i_{2},\cdots,i_{N}| \Pi_{\alpha}|j_{2},\cdots,j_{N}\rangle=0.
\end{equation*}
Based on the above argument, we deduce that some off-diagonal elements of $\Pi_{\alpha}$, shown in $\mathrm{TABLE}$ \ref{tab:2} are zero.
\begin{table}[H]\centering
\caption{Some off-diagonal elements of $\Pi_{\alpha}$ for $N=3n$. Here we apply Block~Zeros~Lemma to the sets $\widetilde{\mathcal{S}_{i}},~\widetilde{\mathcal{S}_{j}}~(i\neq j\in{\{0,1,3\}})$ given by Eq. (\ref{eq:IVS1}).}
\renewcommand\arraystretch{1.8}
\begin{tabular}{c c}
 \hline
 \hline
Sets & ~~~~~~~~Elements \\
\hline
$\widetilde{\mathcal{S}}_{0},\widetilde{\mathcal{S}}_{1}$ & ~~~~~~~~$_{\mathcal{G}^{N-1}_{0}\backslash{\mathbf{\{0\}}}}(\Pi_{\alpha})_{\mathcal{G}^{N-1}_{1}} =0$, ~$_{\mathcal{G}^{N-1}_{0}}(\Pi_{\alpha})_{\mathcal{G}^{N-1}_{2}\backslash{\mathbf{\{1\}}}} =0$, ~$_{\mathcal{G}^{N-1}_{1}\backslash{\mathbf{\{2\}}}}(\Pi_{\alpha})_{\mathcal{G}^{N-1}_{2}} =0$\\
$\widetilde{\mathcal{S}}_{1},\widetilde{\mathcal{S}}_{3}$ & ~~~$_{\{\mathbf{0}\}}(\Pi_{\alpha})_{\mathcal{G}^{N-1}_{1}} =0$, ~~~~~~~$_{\mathcal{G}^{N-1}_{0}}(\Pi_{\alpha})_{\{\mathbf{1}\}}=0$,~~~~~~~~~~$_{\{\mathbf{2}\}}(\Pi_{\alpha})_{\mathcal{G}^{N-1}_{2}}=0$\\
$\widetilde{\mathcal{S}}_{0},\widetilde{\mathcal{S}}_{3}$ & ~~~~~~~~$_{\{\mathbf{0}\}}(\Pi_{\alpha})_{\mathcal{G}^{N-1}_{0}\backslash{\mathbf{\{0\}}}} =0$, ~~~$_{\{\mathbf{1}\}}(\Pi_{\alpha})_{\mathcal{G}^{N-1}_{2}\backslash{\mathbf{\{1\}}}} =0$,~~~$_{\{\mathbf{2}\}}(\Pi_{\alpha})_{\mathcal{G}^{N-1}_{1}\backslash{\mathbf{\{2\}}}} =0$ \\
\hline
\hline
\end{tabular}\label{tab:2}
\end{table}

From TABLE \ref{tab:1}, we know that $\mathbf{0}\in \mathcal{G}^{N-1}_{0}$, which means $\textbf{\emph{i}}_\mathbf{{0}}=(1,0,0,\cdots,0)\in\widetilde{\mathcal{G}^{N}_{1}}$.

For any $\textbf{\emph{j}}=(j_{1}, j_{2},\cdots, j_{N})\in\widetilde{\mathcal{G}^{N}_{1}}$ and $\textbf{\emph{j}}\neq\textbf{\emph{i}}_\mathbf{{0}}$, $j_{1}\neq1$, there is
\begin{equation*}
\langle\textbf{\emph{i}}_\mathbf{{0}}|E|\textbf{\emph{j}}\rangle=\langle\textbf{\emph{i}}_\mathbf{{0}}|\mathbb{I}_{A_{1}}\otimes \Pi_{\alpha}|\textbf{\emph{j}}\rangle=0.
\end{equation*}
If $j_{1}=1$, then we get $\textbf{\emph{j}}\in\{1\}\times \mathcal{G}^{N-1}_{0}$, therefore $(j_{2},\cdots, j_{N})\in\mathcal{G}^{N-1}_{0}$ and $(j_{2},\cdots, j_{N})\neq{\mathbf{0}}$. Noticing that $_{\{\mathbf{0}\}}\Pi_{\mathcal{G}^{N-1}_{0}\backslash{\mathbf{\{0\}}}} =0$ in $\mathrm{TABLE}$ \ref{tab:2}, one obtains
\begin{equation*}
\langle\textbf{\emph{i}}_\mathbf{{0}}|E|\textbf{\emph{j}}\rangle=\langle 1,0,\cdots,0|\mathbb{I}_{A_{1}}\otimes \Pi_{\alpha}|1,j_{2}\cdots,j_{N}\rangle=\langle 0,\cdots,0| \Pi_{\alpha}|j_{2}\cdots,j_{N}\rangle=0.\\
\end{equation*}
Applying Block Trivial Lemma to the set of basis vectors corresponding to $\widetilde{\mathcal{G}^{N}_{1}}$, the set $\widetilde{\mathcal{S}_{1}}$ of states and the vector $|1,0,\cdots,0\rangle$, for any different strings $\textbf{\emph{i}'}=(i'_{1},i'_{2},\cdots,i'_{N})$ and $\textbf{\emph{j}'}=(j'_{1},j'_{2},\cdots,j'_{N})$ belonging to $\widetilde{\mathcal{G}^{N}_{1}}$, we have
\begin{equation*}
\langle \textbf{\emph{i}'}|E|\textbf{\emph{j}'}\rangle=\langle \textbf{\emph{j}'}|E|\textbf{\emph{i}'}\rangle=0,~~~~~\langle \textbf{\emph{i}'}|E|\textbf{\emph{i}'}\rangle=\langle \textbf{\emph{j}'}|E|\textbf{\emph{j}'}\rangle,
\end{equation*}
Which implies that,
\begin{equation*}
\langle i'_{2},\cdots,i'_{N}|\Pi_{\alpha}|i'_{2},\cdots,i'_{N}\rangle=\langle j'_{2},\cdots,j'_{N}|\Pi_{\alpha}|j'_{2},\cdots,j'_{N}\rangle.
\end{equation*}
If $i'_{1}=j'_{1}$, one has
\begin{equation*}
\langle i'_{1},i'_{2},\cdots,i'_{N}|E|j'_{1},j'_{2},\cdots,j'_{N}\rangle=\langle i'_{2},\cdots,i'_{N}|\Pi_{\alpha}|j'_{2},\cdots,j'_{N}\rangle=0.
\end{equation*}
The diagonal elements and some off-diagonal elements of  $\Pi_{\alpha}$ for $N=3n$ are illustrated in TABLE \ref{tab:3}.
\begin{table}[H]\centering
\caption{Diagonal elements and some off-diagonal elements of $\Pi_{\alpha}$ when $N=3n$. Here we apply Block~Trivial~Lemma to the set $\widetilde{\mathcal{S}_{1}}$, the set of basis vectors corresponding to $\widetilde{\mathcal{G}^{N}_{1}}$, and the vector $|1,0,\cdots,0\rangle$.}
\renewcommand\arraystretch{1.6}
\begin{tabular}{c c}
 \hline
 \hline
Subsets of $\widetilde{\mathcal{G}^{N}_{1}}$  & ~~~~~~~~Elements\\
\hline
$\{1\}\times\mathcal{G}^{N-1}_{0}$ & ~~~~~$\langle\textbf{\emph{i}}|\Pi_{\alpha}|\textbf{\emph{j}}\rangle=0$, for $\textbf{\emph{i}}\neq\textbf{\emph{j}}\in\mathcal{G}^{N-1}_{0}$ \\
$\{0\}\times\mathcal{G}^{N-1}_{1}$ & ~~~~~$\langle\textbf{\emph{i}}|\Pi_{\alpha}|\textbf{\emph{j}}\rangle=0$, for $\textbf{\emph{i}}\neq\textbf{\emph{j}}\in\mathcal{G}^{N-1}_{1}$\\
$\{2\}\times\mathcal{G}^{N-1}_{2}$ & ~~~~~$\langle\textbf{\emph{i}}|\Pi_{\alpha}|\textbf{\emph{j}}\rangle=0$, for $\textbf{\emph{i}}\neq\textbf{\emph{j}}\in\mathcal{G}^{N-1}_{2}$\\
$\widetilde{\mathcal{G}^{N}_{1}}$ & ~~~~~$\langle \textbf{\emph{i}}|\Pi_{\alpha}|\textbf{\emph{i}}\rangle=\langle \textbf{\emph{j}}|\Pi_{\alpha}|\textbf{\emph{j}}\rangle$, for $\textbf{\emph{i}}\neq\textbf{\emph{j}}\in\mathbb{Z}^{N-1}_{3}$\\
\hline
\hline
\end{tabular}\label{tab:3}
\end{table}
 Observe the first and second rows of TABLE \ref{tab:2}, the results $_{\mathcal{G}^{N-1}_{0}\backslash{\mathbf{\{0\}}}}(\Pi_{\alpha})_{\mathcal{G}^{N-1}_{1}} =0$ and $_{\{\mathbf{0}\}}(\Pi_{\alpha})_{\mathcal{G}^{N-1}_{1}} =0$ in the first column yield  $_{\mathcal{G}^{N-1}_{0}}(\Pi_{\alpha})_{\mathcal{G}^{N-1}_{1}} =0$. Similarly, we can obtain $_{\mathcal{G}^{N-1}_{0}}(\Pi_{\alpha})_{\mathcal{G}^{N-1}_{2}}=0$ and $_{\mathcal{G}^{N-1}_{1}}(\Pi_{\alpha})_{\mathcal{G}^{N-1}_{2}}=0$ by the second and third columns. Combining the results $\langle\textbf{\emph{i}}|\Pi_{\alpha}|\textbf{\emph{j}}\rangle=0$ for $\textbf{\emph{i}}\neq\textbf{\emph{j}}\in\mathcal{G}^{N-1}_{i}(i\in\{0,1,2\})$ in TABLE \ref{tab:3}, ensures that the off-diagonal elements of $\Pi_{\alpha}$ are all zeros. By the results in last row of TABLE \ref{tab:3}, we obtain  that the diagonal elements of $\Pi_{\alpha}$ are all equal. Thus $\Pi_{\alpha}$ is proportional to the identity for $\alpha=1,2,\cdots$.  Because of the symmetrical structure, we can also show that any $(N-1)$-parties could only perform a trivial OPLM.$\hfill\blacksquare$

For the case of $N=3n+1$ and $N=3n+2$, we give two theorems similar to Theorem 2, what we need to do is to prove that
the set $\widetilde{\mathcal{S}}\backslash\widetilde{\mathcal{S}_{2}}$ has the property of strong nonlocality, which means that any $(N-1)$-parties could only perform a trivial OPLM. We omit the detailed proof but give four tables for the complete analysis, because it is similar to that of Theorem 2.

$\mathbf{Case~II}$: $N=3n+1~(n\geq1)$.

We redefine
\begin{equation}
\begin{aligned}
&\widetilde{\mathcal{G}^{N}_{0}}=\left(\{0\}\times \big(\mathcal{G}^{N-1}_{0}\backslash\{\mathbf{0}\}\big)\right) \bigcup \left(\{2\}\times\mathcal{G}^{N-1}_{1}\right) \bigcup\left(\{1\}\times\mathcal{G}^{N-1}_{2}\right)\label{con:inventoryflow7},\\
&\widetilde{\mathcal{G}^{N}_{1}}=\left(\{1\}\times \big(\mathcal{G}^{N-1}_{0}\backslash\{\mathbf{1}\}\big)\right) \bigcup \left(\{0\}\times \mathcal{G}^{N-1}_{1}\right) \bigcup\left(\{2\}\times \mathcal{G}^{N-1}_{2}\right),\\
&\widetilde{\mathcal{G}^{N}_{2}}=\left(\{2\}\times \mathcal{G}^{N-1}_{0}\right) \bigcup \left(\{1\}\times \mathcal{G}^{N-1}_{1}\right) \bigcup\left(\{0\}\times \mathcal{G}^{N-1}_{2}\right),\\
&\widetilde{\mathcal{G}^{N}_{3}}=\big\{(0,\underbrace{0,\cdots,0}\limits_{N-1}), (1,\underbrace{1,\cdots,1}\limits_{N-1})\big\}.
\end{aligned}
\end{equation}
Here $\mathcal{G}^{N-1}_{i}~(i=0,1,2)$ is given by Eq. (\ref{eq:G}). The set $\mathcal{G}^{N}_{2}$ given by Eq. (\ref{eq:G3}) is renamed as $\widetilde{\mathcal{G}^{N}_{2}}$.
Let $\widetilde{s_{i}}$ be the cardinality of the set $\widetilde{\mathcal{G}^{N}_{i}}$, $i\in \mathbb{Z}_{4}$, we define
\begin{equation}
\begin{aligned}
&\widetilde{\mathcal{S}_{i}}:=\{|\widetilde{\Psi}_{i,k}\rangle \in \mathcal{H}~\big|~k \in \mathbb{Z}_{\widetilde{s_{i}}}, |\widetilde{\Psi}_{i,k}\rangle:=\sum\limits_{\textbf{\emph{j}}\in {\widetilde{\mathcal{G}^{N}_{i}}}}\omega^{k f_{i}(\textbf{\emph{j}})}_{\widetilde{s_{i}}}|\textbf{\emph{j}}\rangle\}.
\end{aligned}\label{eq:IVS2}
\end{equation}
Here $ f_{i}:\widetilde{\mathcal{G}^{N}_{i}}\longrightarrow \mathbb{Z}_{\widetilde{s_{i}}}$ is any fixed bijection and $\omega_{\widetilde{s_{i}}}:=\mathrm{e}^{\frac{2\pi\sqrt{-1}}{\widetilde{s_{i}}}}$.

$\mathbf{Theorem~3}$ In $(\mathbb{C}^{3})^{\otimes N}~(N=3n+1,~n\geq1)$, the set $\widetilde{\mathcal{S}}:=\cup^{3}_{i=0}\widetilde{\mathcal{S}_{i}}$ given by Eq. (\ref{eq:IVS2}) is a strong nonlocal OGEB. The set $\widetilde{\mathcal{S}}\backslash\widetilde{\mathcal{S}_{2}}=\widetilde{\mathcal{S}_{0}}\cup\widetilde{\mathcal{S}_{1}}\cup\widetilde{\mathcal{S}_{3}}$ is a strong nonlocal OGES of size $2\times3^{N-1}$.

$\mathbf{Proof}$. See TABLE \ref{tab:4} and TABLE \ref{tab:5} for the complete analysis. Thus we obtain $\Pi_{\alpha}\varpropto \mathbb{I}$.
\begin{table}[H]\centering
\caption{Some off-diagonal elements of $\Pi_{\alpha}$ when $N=3n+1$. Here we apply Block~Zeros~Lemma to any two different sets $\widetilde{\mathcal{S}_{i}}~(i={0,1,3})$ given by Eq. (\ref{eq:IVS2}).}
\renewcommand\arraystretch{1.6}
\begin{tabular}{c c}
 \hline
 \hline
Sets & ~~~~~~~~Elements \\
\hline
$\widetilde{\mathcal{S}_{0}},\widetilde{\mathcal{S}_{1}}$ & ~~~~~~~~$_{\mathcal{G}^{N-1}_{0}\backslash{\mathbf{\{0\}}}}(\Pi_{\alpha})_{\mathcal{G}^{N-1}_{1}} =0$, ~$_{\mathcal{G}^{N-1}_{0}\backslash{\mathbf{\{1\}}}}(\Pi_{\alpha})_{\mathcal{G}^{N-1}_{2}} =0$, ~$_{\mathcal{G}^{N-1}_{1}}(\Pi_{\alpha})_{\mathcal{G}^{N-1}_{2}} =0$\\
$\widetilde{\mathcal{S}_{1}},\widetilde{\mathcal{S}_{3}}$ & ~~~$_{\{\mathbf{0}\}}(\Pi_{\alpha})_{\mathcal{G}^{N-1}_{1}} =0$, ~~~~~~~$_{\{\mathbf{1}\}}(\Pi_{\alpha})_{\mathcal{G}^{N-1}_{0}\backslash{\mathbf{\{1\}}}}=0$~~~~~~~~~~~~~~~~~~~~\\
$\widetilde{\mathcal{S}_{0}},\widetilde{\mathcal{S}_{3}}$ & ~~~$_{\{\mathbf{0}\}}(\Pi_{\alpha})_{\mathcal{G}^{N-1}_{0}\backslash{\mathbf{\{0\}}}} =0$, ~~~~~~~$_{\{\mathbf{1}\}}(\Pi_{\alpha})_{\mathcal{G}^{N-1}_{2}} =0$~~~~~~~~~~~~~~~~~~~~~~~~~ \\
\hline
\hline
\end{tabular}\label{tab:4}
\end{table}
\begin{table}[H]\centering
\caption{When $N=3n+1$, diagonal elements and some off-diagonal elements of $\Pi_{\alpha}$ are shown. Here we apply Block~Trivial~Lemma to the sets $\widetilde{\mathcal{S}_{1}}$ and $\widetilde{\mathcal{S}_{3}}$ given by Eq. (\ref{eq:IVS2}), the set of base vectors corresponding to $\widetilde{\mathcal{G}^{N}_{1}}$ and $\widetilde{\mathcal{G}^{N}_{3}}$, and the vector $|1,0,\cdots0\rangle$.}
\renewcommand\arraystretch{1.6}
\begin{tabular}{c c c c }
 \hline
 \hline
Subsets of $\widetilde{\mathcal{G}^{N}_{1}}$  & ~~~~~~~~Elements & ~~~~~Set & ~~~~~Elements \\
\hline
$\{1\}\times\mathcal{G}^{N-1}_{0}\backslash\{\mathbf{1}\}$ & ~~~~~ $\langle\textbf{\emph{i}}|\Pi_{\alpha}|\textbf{\emph{j}}\rangle=0$, for $\textbf{\emph{i}}\neq\textbf{\emph{j}}\in\mathcal{G}^{N-1}_{0}\backslash\{\mathbf{1}\}$
&~~~~~$\widetilde{\mathcal{G}^{N}_{1}}$ & ~~~~~~$\langle \textbf{\emph{i}}|\Pi_{\alpha}|\textbf{\emph{i}}\rangle=\langle \textbf{\emph{j}}|\Pi_{\alpha}|\textbf{\emph{j}}\rangle$, for $\textbf{\emph{i}}\neq\textbf{\emph{j}}\in \mathbb{Z}^{N-1}_{3}\backslash\{\mathbf{1}\}$\\
$\{0\}\times\mathcal{G}^{N-1}_{1}$ ~~~~~& $\langle\textbf{\emph{i}}|\Pi_{\alpha}|\textbf{\emph{j}}\rangle=0$,  for $\textbf{\emph{i}}\neq\textbf{\emph{j}}\in\mathcal{G}^{N-1}_{1}$& ~~~~~$\widetilde{\mathcal{G}^{N}_{3}}$ &  ~~~~~$\langle\mathbf{0}|\Pi_{\alpha}|\mathbf{0}\rangle=\langle\mathbf{1}|\Pi_{\alpha}|\mathbf{1}\rangle$\\
$\{2\}\times\mathcal{G}^{N-1}_{2}$ &
$\langle\textbf{\emph{i}}|\Pi_{\alpha}|\textbf{\emph{j}}\rangle=0$, for $\textbf{\emph{i}}\neq\textbf{\emph{j}}\in\mathcal{G}^{N-1}_{2}$\\
\hline
\hline
\end{tabular}\label{tab:5}
\end{table}$\hfill\blacksquare$

$\mathbf{Case~III}$: $N=3n+2~(n\geq1).$

We redefine
\begin{equation}
\begin{aligned}
&\widetilde{\mathcal{G}^{N}_{0}}=\left(\{0\}\times \big(\mathcal{G}^{N-1}_{0}\backslash\{\mathbf{0}\}\big)\right) \bigcup \left(\{2\}\times\mathcal{G}^{N-1}_{1}\right) \bigcup\left(\{1\}\times\mathcal{G}^{N-1}_{2}\right)\label{con:inventoryflow8},\\
&\widetilde{\mathcal{G}^{N}_{1}}=\left(\{1\}\times \mathcal{G}^{N-1}_{0}\right) \bigcup \left(\{0\}\times \mathcal{G}^{N-1}_{1}\right) \bigcup\left(\{2\}\times \big(\mathcal{G}^{N-1}_{2}\backslash\{\mathbf{2}\}\big)\right),\\
&\widetilde{\mathcal{G}^{N}_{2}}=\left(\{2\}\times \mathcal{G}^{N-1}_{0}\right) \bigcup \left(\{1\}\times \mathcal{G}^{N-1}_{1}\right) \bigcup\left(\{0\}\times \mathcal{G}^{N-1}_{2}\right),\\
&\widetilde{\mathcal{G}^{N}_{3}}=\big\{(0,\underbrace{0,\cdots,0}\limits_{N-1}), (2,\underbrace{2,\cdots,2}\limits_{N-1})\big\}.
\end{aligned}
\end{equation}
Here $\mathcal{G}^{N-1}_{i}~(i=0,1,2)$ is given by Eq. (\ref{eq:G}). The set $\mathcal{G}^{N}_{2}$ given by Eq. (\ref{eq:G3}) is renamed as $\widetilde{\mathcal{G}^{N}_{2}}$.
Let $\widetilde{s_{i}}$ be the cardinality of the set $\widetilde{\mathcal{G}^{N}_{i}}$, $i\in \mathbb{Z}_{4}$, we define
\begin{equation}
\begin{aligned}
&\widetilde{\mathcal{S}_{i}}:=\{|\widetilde{\Psi}_{i,k}\rangle \in \mathcal{H}~\big|~k \in \mathbb{Z}_{\widetilde{s_{i}}}, |\widetilde{\Psi}_{i,k}\rangle:=\sum\limits_{\textbf{\emph{j}}\in {\widetilde{\mathcal{G}^{N}_{i}}}}\omega^{k f_{i}(\textbf{\emph{j}})}_{\widetilde{s_{i}}}|\textbf{\emph{j}}\rangle\}.
\end{aligned}\label{eq:IVS3}
\end{equation}
Here $ f_{i}:\widetilde{\mathcal{G}^{N}_{i}}\longrightarrow \mathbb{Z}_{\widetilde{s_{i}}}$ is any fixed bijection and $\omega_{\widetilde{s_{i}}}:=\mathrm{e}^{\frac{2\pi\sqrt{-1}}{\widetilde{s_{i}}}}$.

$\mathbf{Theorem~4}$  In $(\mathbb{C}^{3})^{\otimes N}~(N=3n+2,~n\geq1)$, the set $\widetilde{\mathcal{S}}:=\cup^{3}_{i=0}\widetilde{\mathcal{S}_{i}}$ given by Eq. (\ref{eq:IVS3}) is a strongly nonlocal OGEB. The set $\widetilde{\mathcal{S}}\backslash\widetilde{\mathcal{S}_{2}}=\widetilde{\mathcal{S}_{0}}\cup\widetilde{\mathcal{S}_{1}}\cup\widetilde{\mathcal{S}_{3}}$ is a strongly nonlocal OGES of size $2\times3^{N-1}$.

$\mathbf{Proof}$. The complete analysis please see TABLE \ref{tab:6} and TABLE \ref{tab:7}. Therefore we obtain $\Pi_{\alpha}\varpropto \mathbb{I}$.
\begin{table}[H]\centering
\caption{Some off-diagonal elements of $\Pi_{\alpha}$ when $N=3n+2$. Here we apply Block~Zeros~Lemma to the sets $\widetilde{\mathcal{S}_{i}}~(i={0,1,3})$ given by Eq. (\ref{eq:IVS3}).}
\renewcommand\arraystretch{1.5}
\begin{tabular}{c c}
 \hline
 \hline
Sets & ~~~~~~~~Elements \\
\hline
$\widetilde{\mathcal{S}_{0}},\widetilde{\mathcal{S}_{1}}$ & ~~~~~~~~$_{\mathcal{G}^{N-1}_{0}\backslash{\mathbf{\{0\}}}}(\Pi_{\alpha})_{\mathcal{G}^{N-1}_{1}} =0$, ~$_{\mathcal{G}^{N-1}_{0}}(\Pi_{\alpha})_{\mathcal{G}^{N-1}_{2}} =0$, ~$_{\mathcal{G}^{N-1}_{1}}(\Pi_{\alpha})_{\mathcal{G}^{N-1}_{2}\backslash{\mathbf{\{2\}}}} =0$\\
$\widetilde{\mathcal{S}_{1}},\widetilde{\mathcal{S}_{3}}$ & ~~~$_{\{\mathbf{0}\}}(\Pi_{\alpha})_{\mathcal{G}^{N-1}_{1}} =0$, ~~~~~~~$_{\{\mathbf{2}\}}(\Pi_{\alpha})_{\mathcal{G}^{N-1}_{2}\backslash{\mathbf{\{2\}}}}=0$~~~~~~~~~~~~~~~~~~~~\\
$\widetilde{\mathcal{S}_{0}},\widetilde{\mathcal{S}_{3}}$ & ~~~$_{\{\mathbf{0}\}}(\Pi_{\alpha})_{\mathcal{G}^{N-1}_{0}\backslash{\mathbf{\{0\}}}} =0$, ~~~~~~~$_{\mathcal{G}^{N-1}_{1}}(\Pi_{\alpha})_{\{\mathbf{2}\}}=0$~~~~~~~~~~~~~~~~~~~~~~~~~ \\
\hline
\hline
\end{tabular}\label{tab:6}
\end{table}
\begin{table}[!htp]
\caption{Diagonal elements and some off-diagonal elements of $\Pi_{\alpha}$ when $N=3n+2$. Here one applies Block Trivial Lemma to the sets $\widetilde{\mathcal{S}_{1}}$ and $\widetilde{\mathcal{S}_{3}}$ given by Eq. (\ref{eq:IVS3}), the sets of base vectors corresponding to $\widetilde{\mathcal{G}^{N}_{1}}$ and $\widetilde{\mathcal{G}^{N}_{3}}$, and the vector $|1,0,\cdots0\rangle$.}
\renewcommand\arraystretch{1.5}
\begin{tabular}{c c c c }
 \hline
 \hline
Subsets of $\widetilde{\mathcal{G}^{N}_{1}}$  & ~~~~~~~~Elements & ~~~~~Set & ~~~~~Elements \\
\hline
$\{1\}\times\mathcal{G}^{N-1}_{0}$ & ~~~~~ $\langle\textbf{\emph{i}}|\Pi_{\alpha}|\textbf{\emph{j}}\rangle=0$, for $\textbf{\emph{i}}\neq\textbf{\emph{j}}\in\mathcal{G}^{N-1}_{0}$
&~~~~~$\widetilde{\mathcal{G}^{N}_{1}}$ & ~~~~~~$\langle \textbf{\emph{i}}|\Pi_{\alpha}|\textbf{\emph{i}}\rangle=\langle \textbf{\emph{j}}|\Pi_{\alpha}|\textbf{\emph{j}}\rangle$, for $\textbf{\emph{i}}\neq\textbf{\emph{j}}\in \mathbb{Z}^{N-1}_{3}\backslash\{\mathbf{2}\}$\\
$\{0\}\times\mathcal{G}^{N-1}_{1}$ ~~~~~&$\langle\textbf{\emph{i}}|\Pi_{\alpha}|\textbf{\emph{j}}\rangle=0$, for $\textbf{\emph{i}}\neq\textbf{\emph{j}}\in\mathcal{G}^{N-1}_{1}$ & ~~~~~$\widetilde{\mathcal{G}^{N}_{3}}$ &  ~~~~~$\langle\mathbf{0}|\Pi_{\alpha}|\mathbf{0}\rangle=\langle\mathbf{2}|\Pi_{\alpha}|\mathbf{2}\rangle$\\
$\{2\}\times\mathcal{G}^{N-1}_{2}\backslash\{\mathbf{2}\}$ & ~~~~
$\langle\textbf{\emph{i}}|\Pi_{\alpha}|\textbf{\emph{j}}\rangle=0$, for $\textbf{\emph{i}}\neq\textbf{\emph{j}}\in\mathcal{G}^{N-1}_{2}\backslash\{\mathbf{2}\}$\\
\hline
\hline
\end{tabular}\label{tab:7}
\end{table}$\hfill\blacksquare$

We construct strongly nonlocal OGESs containing $2\times3^{N-1}$ states in $(\mathbb{C}^{3})^{\otimes N}$, which is $3^{N-1}-2^{N}+1$ fewer than the OGESs presented in Ref. \cite{M. S. Li1} and $3^{N-1}-1$  fewer than the OPSs in Ref. \cite{Y. He,H. Q. Zhou2}. Moreover, it is also $3^{N-1}-2^{N}$ fewer than the UPBs in Ref. \cite{Y. He}. It should be pointed out that our OGESs are also of the strongest nonlocality. A set of orthogonal states is said to have the property of the strongest nonlocality \cite{Y. L. Wang} if only a trivial orthogonality-preserving POVM can be performed for each bipartition of the subsystems. As a consequence, we successfully show that there do exist smaller size of strongest nonlocal OGESs in an $N$-qutrit system.

\section{Conclusion}\label{Q4}
In this work, we constructed OGESs and OGEBs with strong nonlocality in $(\mathbb{C}^{3})^{\otimes N}~(N\geq3)$, which positively answer the open question in Ref. \cite{S. Halder2} of "whether one can find orthogonal entangled bases that are locally irreducible in all bipartitions". Furthermore, in an $N$-qutrit system, the strongly nonlocal OGESs in our construction has much smaller size than that of strongly nonlocal OGESs in Ref. \cite{M. S. Li1} and strongly nonlocal OPSs in Refs. \cite{Y. He,H. Q. Zhou2}. Thus, this work is also a answer to the question in Ref. \cite{Y. L. Wang}, "can we construct some smaller set that has the property of the strongest nonlocality via the OGES than the OPS". Our result could also be helpful in better understanding the structure of "local irreducibility in all bipartitions " of entangled states.
\section{Acknowledgments}
This work was supported by the National Natural Science Foundation of China under Grants No. 12071110 and No. 62271189, the Hebei Central Guidance on Local Science and Technology Development Foundation of China under Grant No. 236Z7604G, and by the Science and Technology Project of Hebei Education Department under Grant No. ZD2021066.


\begin{thebibliography}{99}

\bibitem{N. Brunner} N. Brunner, D. Cavalcanti, S. Pironio, V. Scarani, and S. Wehner, Bell nonlocality, \href{https://journals.aps.org/rmp/abstract/10.1103/RevModPhys.86.419} {Rev. Mod. Phys. \textbf{86}, 419 (2014)}; Erratum, \href{https://journals.aps.org/rmp/abstract/10.1103/RevModPhys.86.839} {Rev. Mod. Phys. \textbf{86}, 839 (2014)}.

\bibitem{R. Horodecki} R. Horodecki, P. Horodecki, M. Horodecki, and K. Horodecki, Quantum entanglement, \href{https://journals.aps.org/rmp/abstract/10.1103/RevModPhys.81.865} {Rev. Mod. Phys. \textbf{81}, 865 (2009)}.

\bibitem{T. Gao} T. Gao, F. L. Yan, and S. J. van Enk, Permutationally invariant part of a density matrix and nonseparability of
$N$-qubit states, \href{https://journals.aps.org/prl/abstract/10.1103/PhysRevLett.112.180501} {Phys. Rev. Lett. \textbf{112}, 180501 (2014)}.

\bibitem{J. S. Bell} J. S. Bell, On the Einstein Podolsky Rosen paradox, \href{https://journals.aps.org/ppf/abstract/10.1103/PhysicsPhysiqueFizika.1.195} {Phys. Phys. Fiz. \textbf{1}, 195 (1964)}; On the problem of hidden variables in quantum mechanics, \href{https://journals.aps.org/rmp/abstract/10.1103/RevModPhys.38.447} {Rev. Mod. Phys. \textbf{38}, 447 (1966)}.

\bibitem{J. F. Clauser} J. F. Clauser, M. A. Horne, A. Shimony, and R. A. Holt, Proposed experiment to test local hidden-variable theories, \href{https://journals.aps.org/prl/abstract/10.1103/PhysRevLett.23.880} {Phys. Rev. Lett. \textbf{23}, 880 (1970)}.

\bibitem{S. J. Freedman} S. J. Freedman and J. F. Clauser, Experimental test of local hidden-variable theories, \href{https://journals.aps.org/prl/abstract/10.1103/PhysRevLett.28.938} {Phys. Rev. Lett. \textbf{28}, 938 (1972)}.

\bibitem{A. Aspect} A. Aspect, P. Grangier, and G. Roger, Experimental tests of realistic local theories via Bell's theorem, \href{https://journals.aps.org/prl/abstract/10.1103/PhysRevLett.47.460} {Phys. Rev. Lett. \textbf{47}, 460 (1981)}.

\bibitem{Z. Q. Chen} Z. Q. Chen, Bell-Klyshko inequalities to characterize maximally entangled states of $n$ qubits, \href{https://journals.aps.org/prl/abstract/10.1103/PhysRevLett.93.110403} {Phys. Rev. Lett. \textbf{93}, 110403 (2004)}.

\bibitem{F. L. Yan} F. L. Yan, T. Gao, and E. Chitambar, Two local observables are sufficient to characterize maximally entangled states of $N$ qubits, \href{https://journals.aps.org/pra/abstract/10.1103/PhysRevA.83.022319} {Phys. Rev. A \textbf{83}, 022319 (2011)}.

\bibitem{D. Ding1} D. Ding, Y. Q. He, F. L. Yan, and T. Gao, Quantum nonlocality of generic family of four-qubit entangled pure states, \href{https://iopscience.iop.org/article/10.1088/1674-1056/24/7/070301} {Chin. Phys. B \textbf{24}, 070301 (2015)}.

\bibitem{H. X. Meng} H. X. Meng, J. Zhou, Z. P. Xu, H. Y. Su, T. Gao, F. L. Yan, and J. L. Chen, Hardy's paradox for multisetting high-dimensional systems, \href{https://journals.aps.org/pra/abstract/10.1103/PhysRevA.98.062103} {Phys. Rev. A \textbf{98}, 062103 (2018)}.

\bibitem{D. Ding2} D. Ding, Y. Q. He, F. L. Yan, and T. Gao, Entanglement measure and quantum violation of Bell-type inequality, \href{https://link.springer.com/article/10.1007/s10773-016-3048-1} {Int. J. Theor. Phys. \textbf{55}, 4231 (2016)}.


\bibitem{C. H. Bennett1} C. H. Bennett, G. Brassard, C. Cr$\acute{\mathrm{e}}$peau, R. Jozsa, A. Peres, and W. K. Wootters, Teleporting an unknown quantum state via dual classical and Einstein-Podolsky-Rosen channels, \href{https://journals.aps.org/prl/abstract/10.1103/PhysRevLett.70.1895} {Phys. Rev. Lett. \textbf{70}, 1895 (1993)}.

\bibitem{T. Gao2} T. Gao, F. L. Yan, and Z. X. Wang, Controlled quantum teleportation and secure direct communication, \href{https://iopscience.iop.org/article/10.1088/1009-1963/14/5/006} {Chin. Phys. \textbf{14}, 893 (2005)}.

\bibitem{T. Gao3} T. Gao, F. L. Yan, and Y. C. Li, Optimal controlled teleportation, \href{https://iopscience.iop.org/article/10.1209/0295-5075/84/50001} {Europhys. Lett. \textbf{84}, 50001 (2008)}.

\bibitem{A. K. Ekert} A. K. Ekert, Quantum cryptography based on Bell's theorem, \href{https://journals.aps.org/prl/abstract/10.1103/PhysRevLett.67.661} {Phys. Rev. Lett. \textbf{67}, 661 (1991)}.

\bibitem{C. H. Bennett2} C. H. Bennett, G. Brassard, and N. D. Mermin, Quantum cryptography without Bell's theorem, \href{https://journals.aps.org/prl/abstract/10.1103/PhysRevLett.68.557} {Phys. Rev. Lett. \textbf{68}, 557 (1992)}.

\bibitem{H. K. Lo} H. K. Lo and H. F. Chau, Unconditional security of quantum key distribution over arbitrarily long distances, \href{https://www.science.org/doi/10.1126/science.283.5410.2050} {Science \textbf{283}, 2050 (1999)}.

\bibitem{S. Perseguers} S. Perseguers, G. J. Lapeyre Jr., D. Cavalcanti, M. Lewenstein, and A. Ac\'{\i}n, Distribution of entanglement in large-scale quantum networks, \href{https://iopscience.iop.org/article/10.1088/0034-4885/76/9/096001} {Rep. Prog. Phys. \textbf{76}, 096001 (2013)}.

\bibitem{E. Chitambar} E. Chitambar, D. Leung, L. Man\v{c}inska, M. Ozols, and A. Winter, Everything you always wanted to know about LOCC (But were afraid to ask) \href{https://link.springer.com/article/10.1007/s00220-014-1953-9} {Commun. Math. Phys. \textbf{328}, 303 (2014)}.

\bibitem{C. H. Bennett} C. H. Bennett, D. P. DiVincenzo, C. A. Fuchs, T. Mor, E. Rains, P. W. Shor, J. A. Smolin, and W. K. Wootters, Quantum nonlocality without entanglement, \href{https://journals.aps.org/pra/abstract/10.1103/PhysRevA.59.1070} {Phys. Rev. A \textbf{59}, 1070 (1999)}.

\bibitem{J. Walgate1} J. Walgate, A. J. Short, L. Hardy, and V. Vedral, Local distinguishability of multipartite orthogonal quantum states, \href{https://journals.aps.org/prl/abstract/10.1103/PhysRevLett.85.4972} {Phys. Rev. Lett. \textbf{85}, 4972 (2000)}.

\bibitem{N. Yu1} N. Yu, R. Duan, and M. Ying, Any $2\otimes n$ subspace is locally distinguishable, \href{https://journals.aps.org/pra/abstract/10.1103/PhysRevA.84.012304} {Phys. Rev. A \textbf{84}, 012304 (2011)}.

\bibitem{J. Niset} J. Niset and N. J. Cerf, Multipartite nonlocality without entanglement in many dimensions, \href{https://journals.aps.org/pra/abstract/10.1103/PhysRevA.74.052103} {Phys. Rev. A \textbf{74}, 052103 (2006)}.

\bibitem{Y. Feng} Y. Feng and Y. Shi, Characterizing locally indistinguishable orthogonal product states, \href{https://ieeexplore.ieee.org/document/4957660} {IEEE Trans. Inf. Theory \textbf{55}, 2799 (2009)}.

\bibitem{Z. C. Zhang1} Z. C. Zhang, F. Gao, G. J. Tian, T. Q. Cao, and Q. Y. Wen,
    Nonlocality of orthogonal product basis quantum states, \href{https://journals.aps.org/pra/abstract/10.1103/PhysRevA.90.022313} {Phys. Rev. A \textbf{90}, 022313 (2014)}.

\bibitem{Y. L. Wang1} Y. L. Wang, M. S. Li, Z. J. Zheng, and S. M. Fei, Nonlocality of orthogonal product-basis quantum states, \href{https://journals.aps.org/pra/abstract/10.1103/PhysRevA.92.032313} {Phys. Rev. A \textbf{92}, 032313 (2015)}.

\bibitem{Y. L. Wang2} Y. L. Wang, M. S. Li, Z. J. Zheng, and S. M. Fei, The local indistinguishability of multipartite product states, \href{https://link.springer.com/article/10.1007/s11128-016-1477-7} {Quantum Inf. Process. \textbf{16}, 5 (2017)}.

\bibitem{S. Halder1} S. Halder, Several nonlocal sets of multipartite pure orthogonal product states, \href{https://journals.aps.org/pra/abstract/10.1103/PhysRevA.98.022303} {Phys. Rev. A \textbf{98}, 022303 (2018)}.

\bibitem{S. Ghosh1} S. Ghosh, G. Kar, A. Roy, A. Sen (De), and U. Sen, Distinguishability of Bell states, \href{https://journals.aps.org/prl/abstract/10.1103/PhysRevLett.87.277902} {Phys. Rev. Lett. \textbf{87}, 277902 (2001)}.

\bibitem{S. Ghosh2} S. Ghosh, G. Kar, A. Roy, and D. Sarkar, Distinguishability of maximally entangled states, \href{https://journals.aps.org/pra/abstract/10.1103/PhysRevA.70.022304} {Phys. Rev. A \textbf{70}, 022304 (2004)}.

\bibitem{H. Fan} H. Fan, Distinguishability and indistinguishability by local operations and classical communication, \href{https://journals.aps.org/prl/abstract/10.1103/PhysRevLett.92.177905} {Phys. Rev. Lett. \textbf{92}, 177905 (2004)}.

\bibitem{N. Yu2} N. Yu, R. Duan, and M. Ying, Four locally indistinguishable ququad-ququad orthogonal maximally entangled states, \href{https://journals.aps.org/prl/abstract/10.1103/PhysRevLett.109.020506} {Phys. Rev. Lett. \textbf{109}, 020506 (2012)}.

\bibitem{B. M. Terha} B. M. Terhal, D. P. DiVincenzo, and D. W. Leung, Hiding bits in Bell states, \href{https://journals.aps.org/prl/abstract/10.1103/PhysRevLett.86.5807} {Phys. Rev. Lett. \textbf{86}, 5807 (2001)}.

\bibitem{D. P. DiVincenzo} D. P. DiVincenzo, D. W. Leung, and B. M. Terhal, Quantum data hiding, \href{https://ieeexplore.ieee.org/document/985948} {IEEE Trans. Inf. Theory \textbf{48}, 580 (2002)}.

\bibitem{R. Rahaman} R. Rahaman and M. G. Parker, Quantum scheme for secret sharing based on local distinguishability, \href{https://journals.aps.org/pra/abstract/10.1103/PhysRevA.91.022330} {Phys. Rev. A \textbf{91}, 022330 (2015)}.

\bibitem{J. Wang} J. Wang, L. Li, H. Peng, and Y. Yang, Quantum-secret-sharing scheme based on local distinguishability of orthogonal multiqudit entangled states, \href{https://journals.aps.org/pra/abstract/10.1103/PhysRevA.95.022320} {Phys. Rev. A \textbf{95}, 022320 (2017)}.

\bibitem{S. Halder2} S. Halder, M. Banik, S. Agrawal, and S. Bandyopadhyay, Strong quantum nonlocality without entanglement, \href{https://journals.aps.org/prl/abstract/10.1103/PhysRevLett.122.040403} {Phy. Rev. Lett. \textbf{122}, 040403 (2019)}.

\bibitem{Z. C. Zhang2} Z. C. Zhang and X. Zhang, Strong quantum nonlocality in multipartite quantum systems, \href{https://journals.aps.org/pra/abstract/10.1103/PhysRevA.99.062108} {Phys. Rev. A \textbf{99}, 062108 (2019)}.

\bibitem{S. Rout} S. Rout, A. G. Maity, A. Mukherjee, S. Halder, and M. Banik, Genuinely nonlocal product bases: Classification and entanglement-assisted discrimination, \href{https://journals.aps.org/pra/abstract/10.1103/PhysRevA.100.032321} {Phys. Rev. A \textbf{100}, 032321 (2019)}.

\bibitem{P. Yuan} P. Yuan, G. J. Tian, and X. M. Sun, Strong quantum nonlocality without entanglement in multipartite quantum systems, \href{https://journals.aps.org/pra/abstract/10.1103/PhysRevA.102.042228} {Phys. Rev. A \textbf{102}, 042228 (2020)}.


\bibitem{F. Shi2} F. Shi, M. S. Li, M. Y. Hu, L. Chen, M. H. Yung, Y. L. Wang, and X. D. Zhang, Strongly nonlocal unextendible product bases do exist, \href{https://quantum-journal.org/papers/q-2022-01-05-619} {Quantum \textbf{6}, 619 (2022)}; Strong quantum nonlocality from hypercubes, \href{https://arxiv.org/abs/2110.08461} {arXiv: 2110.08461}.

\bibitem{B. C. Che} B. C. Che, Z. Dou, M. Lei, and Y. X. Yang, Strong nonlocal sets of UPB, \href{https://arxiv.org/abs/2106.08699v2} {arXiv: 2106.08699v2}.

\bibitem{H. Q. Zhou1} H. Q. Zhou, T. Gao, and F. L. Yan, Orthogonal product sets with strong quantum nonlocality on a plane structure, \href{https://journals.aps.org/pra/abstract/10.1103/PhysRevA.106.052209} {Phys. Rev. A \textbf{106}, 052209 (2022)}.

\bibitem{Y. He} Y. Y. He, F. Shi, and X. D. Zhang, Strong quantum nonlocality and unextendibility without entanglement in $N$-partite systems with odd $N$, \href{https://arxiv.org/abs/2203.14503v2} {arXiv: 2203.14503}.

\bibitem{H. Q. Zhou2} H. Q. Zhou, T. Gao, and F. L. Yan, Strong quantum nonlocality without entanglement in an $n$-partite system with even $n$, \href{https://journals.aps.org/pra/abstract/10.1103/PhysRevA.107.042214} {Phys. Rev. A \textbf{107}, 042214 (2023)}.


\bibitem{F. Shi3} F. Shi, M. Y. Hu, L. Chen, and X. D. Zhang, Strong quantum nonlocality with entanglement, \href{https://journals.aps.org/pra/abstract/10.1103/PhysRevA.102.042202} {Phys. Rev. A \textbf{102}, 042202 (2020)}.

\bibitem{Y. L. Wang} Y. L. Wang, M. S. Li, and M. H. Yung, Graph-connectivity-based strong quantum nonlocality with genuine entanglement, \href{https://journals.aps.org/pra/abstract/10.1103/PhysRevA.104.012424} {Phys. Rev. A \textbf{104}, 012424 (2021)}.
\bibitem{F. Shi1} F. Shi, Z. Ye, L. Chen, and X. Zhang, Strong quantum nonlocality in $N$-partite systems, \href{https://journals.aps.org/pra/abstract/10.1103/PhysRevA.105.022209} {Phys. Rev. A \textbf{105}, 022209 (2022)}.

\bibitem{M. S. Li1} M. S. Li and Y. L. Wang, Bounds on the smallest sets of quantum states with special quantum nonlocality, \href{https://arxiv.org/abs/2202.09034} {arXiv: 2202.09034}.

\bibitem{P. Bej1} P. Bej and S. Halder, Unextendible product bases, bound entangled states, and the range criterion, \href{https://www.sciencedirect.com/science/article/pii/S0375960120308598} {Phys. Lett. A \textbf{386}, 126992 (2021)}.

\bibitem{J. Walgate2} J. Walgate and L. Hardy, Nonlocality, asymmetry, and distinguishing bipartite states,
    \href{https://journals.aps.org/prl/abstract/10.1103/PhysRevLett.89.147901} {Phys. Rev. Lett. \textbf{89}, 147901 (2002)}.

\bibitem{D. Goyeneche} D. Goyeneche and K. \.{Z}yczkowski, Genuinely multipartite entangled states and orthogonal arrays, \href{https://journals.aps.org/pra/abstract/10.1103/PhysRevA.90.022316} {Phys. Rev. A \textbf{90}, 022316 (2014)}.

\bibitem{W. M. Shang} W. M. Shang, X. Y. Fan, F. L. Zhang, and J. L. Chen, Quantum information masking of an arbitrary unknown state can be realized in the multipartite lower-dimensional systems, \href{https://iopscience.iop.org/article/10.1088/1402-4896/acb7ad} {Phys. Scr. \textbf{98}, 035102 (2023)}.




\end{thebibliography}
\end{document}